\documentclass[16pt,a4paper]{extarticle}

\usepackage[utf8]{inputenc}
\usepackage[T1]{fontenc}
\usepackage{amsmath,amssymb,amsfonts}
\usepackage{geometry}
\usepackage{graphicx}
\usepackage{booktabs}
\usepackage{array}
\usepackage{hyperref}
\usepackage{cite}
\usepackage{enumitem}
\usepackage{xcolor}

\title{Orientational average in nuclear shape imaging:\\
densities, correlations, and eccentricity in Ne-20}
\author{%
\normalsize
Jian Deng\,$^{1}$\thanks{jdeng@sdu.edu.cn},
Hao-jie Xu\,$^{2}$,
and Qun Wang\,$^{3,4}$\thanks{qunwang@ustc.edu.cn}\\[0.6em]
\parbox{\dimexpr\textwidth-2\tabcolsep\relax}{\centering\small
$^{1}$\,Key Laboratory of Particle Physics and Particle Irradiation (MOE),
Institute of Frontier and Interdisciplinary Science,
Shandong University, Qingdao, Shandong 266237, China\\[0.25em]
$^{2}$\,School of Science, Huzhou Normal University,
Huzhou, Zhejiang 313000, China\\[0.25em]
$^{3}$\,Department of Modern Physics and Anhui Center for Fundamental Sciences in Theoretical Physics,
University of Science and Technology of China, Hefei, Anhui 230026, China\\[0.25em]
$^{4}$\,School of Mechanics and Physics, Anhui University of Science and Technology,
Huainan, Anhui 232001, China}}
\date{\today}

\begin{document}

\maketitle

%====================================================================
\begin{abstract}
The orientational treatment of deformed nuclei in high-energy nuclear collisions have two ways, a rigid rotor and a wave-function rotor, in which nuclear densities in laboratory frame are constructed differently. 
We calculate the one- and two-body densities (including Pauli blocking effects), body-axis contour multipoles, and geometric eccentricity variances under a leading-order ultracentral mapping for Ne-20 in a deformed harmonic-oscillator model in the rigid and wave-function rotors. At fixed deformation parameter $\kappa$ in the model, the two constructions can give different laboratory radial moments and pair correlations. 
The body-axis transition-kernel contours in the wave-function rotor are generally more compact along the symmetry axis, and $\beta_{20}$, $\beta_{40}$, $\langle\varepsilon_2^2\rangle$ and $\langle\varepsilon_4^2\rangle$ lie below the rigid-rotor results in a wide range of $\kappa$. 
The relations among these quantities in each rotor are nevertheless similar in most scanned parameter space: 
$\beta_{40}$ and $\langle\varepsilon_2^2\rangle$ depend linearly on $\beta_{20}^{2}$, while the hexadecapole baseline-subtracted variance follows an approximate quadratic trend in the quadrupole one. 
The method can be extended to seek more sensitive quantities for rotational effects in the wave-function rotor. 
\end{abstract}

%====================================================================
\section{Introduction}
\label{sec:intro}
High-energy nucleus-nucleus collisions provide a femtoscopic snapshot of the nucleon spatial distribution inside colliding nuclei.
The collective anisotropic flow of hadrons produced in collisions is, to a good approximation, a linear response to the spatial eccentricity $\varepsilon_n$ of nucleons inside colliding nuclei just before collisions. So flow harmonics $v_n$ of hadrons can be used to infer the geometry of the nuclear overlap at the initial stage of collisions \cite{Heinz:2013th}. 
In this imaging program, a deformed nucleus is treated as a rigid rotor \cite{Bohr:1975zz}, an anisotropic density, often described by a deformed Woods--Saxon profile, is oriented at random in the laboratory frame, and the overlap of colliding nuclei determines $\varepsilon_n$ and then $v_n$ after hydrodynamic evolution. 
A lot of progress has been made in mapping the initial-state nuclear deformation to final-state anisotropic flows, nuclear deformation was introduced into the Glauber model as part of initial conditions of collisions \cite{Shou:2014eya}, flow harmonics were used to constrain the quadrupole deformation \cite{Giacalone:2019pca,Giacalone:2020awm}, and the consistency of the nuclear deformation parameter $\beta$ with the elliptic flow $v_2$ was examined \cite{Giacalone:2021udy}. The imprint of the nuclear structure was separated from the response of the medium by comparing nuclei of similar mass and different ground-state geometry \cite{Li:2019kkh,Jia:2022qgl}. 
The measured flow is thereby used, nucleus by nucleus, to determine nuclear shape parameters across the nuclide chart, complementing the nuclear structure method \cite{Jia:2022ozr}.
Collective flows in $^{96}$Ru+$^{96}$Ru and $^{96}$Zr+$^{96}$Zr isobar collisions \cite{STAR:2021mii} have been interpreted as the result of quadrupole and octupole deformation of colliding nuclei \cite{Zhang:2021kxj}. 
In collisions of two U-238 nuclei \cite{STAR:2015mki}, the quadrupole deformation and triaxiality of U-238 can be extracted relative to Au+Au collisions \cite{STAR:2024wgy}, and its hexadecapole deformation can also be specified \cite{Ryssens:2023fkv,Xu:2024bdh}. 
Anisotropic flows in Xe+Xe collisions have been measured by ALICE and CMS collaborations \cite{ALICE:2018lao,CMS:2019cyz} and used to establish the triaxiality of Xe-129 \cite{Bally:2021qys}.
Correlations between flow and transverse momentum in Xe+Xe and Pb+Pb collisions at the LHC have been measured by ATLAS Collaboration as a probe to nuclear deformation \cite{ATLAS:2022dov,Jia:2021wbq}.

The collective flows are more sensitive to the initial geometry in smaller collision systems \cite{Song:2017wtw,Zhao:2017rgg,Zhao:2020wcd}.
Collisons of protons, deuterons, and He-3 on gold nuclei correspond to three distinct geometries \cite{PHENIX:2018lia}.
In high-multiplicity events in pp and p+Pb collisions at the LHC energy, event-by-event fluctuations dominate the initial geometry \cite{Nagle:2018nvi}. 
Although the ground states of O-16 and Ne-20 are in the same $J^\pi=0^+$ state which are spherically symmetric in the laboratory frame, the wave functions are superposition of states with intrinsic shapes over different orientations \cite{Ring:1980}. 
According to lattice effective-field-theory and projected generator-coordinate calculations \cite{Lee:2025req,Frosini:2021sxj}, O-16 is a tetrahedral arrangement of $\alpha$ clusters \cite{YuanyuanWang:2024sgp,Hu:2025eid,Zhao:2024feh} and Ne-20 is a bowling pin \cite{Liu:2025zsi,Li:2025hae}. 
Wave functions from nuclear structure calculations have been used to predict the collective flow of Ne-20 in small systems \cite{Giacalone:2024luz}. The elliptic and triangular flow in $^{16}$O+$^{16}$O and $^{20}$Ne+$^{20}$Ne collisions have been measured by ALICE, ATLAS, and CMS collaboration for nuclear structure studies \cite{ALICE:2025luc,ATLAS:2025nnt,CMS:2025tga}.

The nuclear densities in laboratory frame in nuclear collisions can be constructed in different ways.
Some of us showed that a hierarchy of $n$-body effective densities of even-even nuclei can be constructed directly from the eikonal scattering matrix based on a nonorthogonal Generator Coordinate Method construction of rotationally invariant ground states of even-even nuclei \cite{Xu:2026llz}. 
Within a linearized description of initial-state fluctuations in nuclear collisions, relations between N-particle correlations in the final states of collisions and N-nucleon density distributions in the colliding nuclei can be constructed \cite{Giacalone:2023hwk}. The eccentricity variance in ultracentral collisions can be identified by the expectation value of a two-body operator, with a Gaussian rigid rotor as the benchmark \cite{Duguet:2025hwi}.
A classical random rotation of an axially deformed density leads to a characteristic angular modulation of the laboratory two-body density \cite{Blaizot:2025scr}, and ab initio calculations of that modulation give a quadrupole component in Ne-20 and a triangular component in O-16 \cite{Blaizot:2025bfu}. 
The two average procedures can lead to different laboratory pair correlations even from the same intrinsic state. 
In a Glauber model with a quantum superposition of orientations the eccentricity in central Ne-Ne and Ne-Pb collisions can be lowered \cite{Ke:2025tyv}. Projected wave functions have been used to image two-body correlations in Ne-20 \cite{Bofos:2026huw}. The validity of classical rigid rotor is assessed systematically for several even-even deformed nuclei across the nuclear chart by comparing the quantum quadrupole rotor with its classical rigid-rotor limit in Projected Hartree-Fock-Bogoliubov calculations \cite{Bofos:2026nmg}.

In the Glauber model, the eikonal phase for small-angle multiple scatterings at high energy can be written as an integral over positions of nucleons in projectile ($P$) and target ($T$) nuclei for the overlap of nuclear many-body densities~\cite{Miller:2007ri},
\begin{equation}
\label{eq:glauber-phase}
e^{i\chi(\mathbf{b})}
=\!\int\! \prod_k d^3\mathbf{r}_k^{P}
\prod_l d^3\mathbf{r}_l^{T}\,
\rho_{\rm lab}^P\bigl(\tfrac{\mathbf{b}}{2},\{\mathbf{r}^{P}\}\bigr)\,
\rho_{\rm lab}^T\bigl(-\tfrac{\mathbf{b}}{2},\{\mathbf{r}^{T}\}\bigr)
\prod_{i,j}\bigl[1-\Gamma_{NN}(\mathbf{b}_{ij})\bigr],
\end{equation}
where $\mathbf{b}$ is the impact parameter, $\mathbf{b}_{ij}=(\mathbf{r}_i^{P}-\mathbf{r}_j^{T})_\perp$, and $\Gamma_{NN}$ is a standard Gaussian nucleon-nucleon profile. 
Expanding the product of $(1-\Gamma_{NN})$ shows that the linear term involves the laboratory one-body densities $\rho_{1,\rm lab}$, while terms involving two nucleons from the same nucleus require its laboratory two-body density $\rho_{2,\rm lab}$ and cannot be reconstructed from $\rho_{1,\rm lab}\rho_{1,\rm lab}$ alone~\cite{Alvioli:2009ab,Broniowski:2010jd,Blaizot:2014wba}.
Thus any quantitative Glauber calculation for deformed nuclei rests on a definite prescription for $\rho_{1,\rm lab}$ and $\rho_{2,\rm lab}$.

Two natural prescriptions differ by where the orientational average is performed: on the intrinsic probability density (rigid rotor), or on the many-body wave function before densities are formed (wave-function rotor).  
The purpose of this work is to examine the difference between two prescriptions quantitatively in an analytically solvable model for Ne-20. We use the leading-order ultracentral eccentricity relation of Ref.~\cite{Duguet:2025hwi} as a common mapping of their laboratory densities to initial geometry, and a collision and medium-response calculation is needed to predict measured flow.

The paper is organized as follows. In Section \ref{sec:formal}, the rigid-rotor and wave-function-rotor average procedures at the many-body level are defined, and the laboratory one- and two-body densities entering Eq. \eqref{eq:glauber-phase} from Slater determinants are evaluated. Both rigid-rotor and wave-function-rotor constructions are realized in Section \ref{sec:nilsson} using a Nilsson model for Ne-20 in which nuclear densities and surface multipole deformation from isodensity contours are obtained. 
Ultracentral eccentricity as a correlation-sensitive benchmark is calculated in Section \ref{sec:ecc} with comparison of results for rigid and wave-function rotors. We make a summary and outlook in Section \ref{sec:summary}. 

%====================================================================
\section{Rigid rotor versus wave-function rotor}
\label{sec:formal}

Having identified the laboratory nucleon distribution functions as the microscopic input to Eq.~\eqref{eq:glauber-phase}, we now define the two averaging procedures that generate them from a given intrinsic many-body state $\Psi_{\rm intr}$. We will use superscripts \(\rm rg\) and \(\rm wf\) to label the rigid-rotor and wave-function-rotor densities, respectively.

The internal state of a nucleus with \(A\) nucleons is described by the many-body wave function $\Psi_{\rm intr}(x_1,\dots,x_A)$, where the collective coordinate \(x_i=\{\mathbf{r}_i,s_i,\tau_i\}\) comprises the spatial position, spin, and isospin of the \(i\)-th nucleon, with measure $dx_i=d^3\mathbf{r}_i\sum_{s_i,\tau_i=\pm 1/2}$.

In the rigid-rotor picture, the intrinsic many-body probability density can be defined as 
\begin{equation}
\label{eq:rho-c-intr}
\rho^{\rm rg}_{\rm intr}(x_1,\dots,x_A)
=\bigl|\Psi_{\rm intr}(x_1,\dots,x_A)\bigr|^2.
\end{equation}
Here $\rho^{\rm rg}_{\rm intr}$ may not be spherically symmetric, but laboratory isotropy is restored by averaging the intrinsic density over all spatial orientations,
\begin{equation}
\label{eq:rho-c-lab}
\rho^{\rm rg}_{\rm lab}(x_1,\dots,x_A)
=\frac{1}{8\pi^2}\int d\Omega\,\hat{R}(\Omega)\,\rho^{\rm rg}_{\rm intr}(x_1,\dots,x_A), 
\end{equation}
where $\hat{R}(\Omega)$ is the spatial rotation operator parametrized by Euler angles $\Omega=(\phi,\theta,\gamma)$, with factorization $\hat{R}(\Omega)=\hat{R}_z(\phi)\hat{R}_y(\theta)\hat{R}_z(\gamma)$ and invariant measure $d\Omega$; it acts on functions to its right by rotating the spatial arguments, $\hat{R}(\Omega)f(\{x_i\})=f(\{R^{-1}(\Omega)x_i\})$. 
Because $\Psi_{\rm intr}$ is normalized and the Haar average merely reshuffles the arguments of an already normalized scalar density, both the intrinsic and laboratory rigid-rotor densities inherit unit normalization without any additional factor:
\begin{equation}
\label{eq:rho-c-norm}
\int\prod_{i=1}^A dx_i\,\rho^{\rm rg}_{\rm intr}(x_1,\dots,x_A)
=\int\prod_{i=1}^A dx_i\,\rho^{\rm rg}_{\rm lab}(x_1,\dots,x_A) =1.
\end{equation}
The full laboratory distribution is rotationally invariant. Its one-body marginal is spherical, while its two-body marginal can retain dependence on relative angles. Marginalizing Eq.~\eqref{eq:rho-c-lab} to one body reproduces the usual orientational average of an intrinsic one-body density.

In the wave-function-rotor picture, the laboratory state is required to be an eigenstate of angular momentum, so the rotational average is performed on the wave function instead of on the density, preserving the relative phase between bra and ket state: 
\begin{equation}
\label{eq:Psi-lab}
\Psi_{\rm lab}(x_1,\dots,x_A)
=\frac{1}{8\pi^2\mathcal{N}_A}\int d\Omega\,\hat{R}(\Omega)\,
\Psi_{\rm intr}(x_1,\dots,x_A)\;.
\end{equation}
Unlike Eq.~\eqref{eq:rho-c-norm}, this coherent superposition of differently oriented components is not normalized by the Haar measure alone. The constant $\mathcal{N}_A$ must be introduced to restore the normalization of $\Psi_{\rm lab}$, which 
is fixed by $\mathcal{N}_A^2=\langle\Psi_{\rm intr}|P_0|\Psi_{\rm intr}\rangle$ with $P_0=(8\pi^2)^{-1}\int d\Omega\,\hat R(\Omega)$. The laboratory many-body density is then 
\begin{equation}
\label{eq:rho-q-lab}
\rho^{\rm wf}_{\rm lab}(x_1,\dots,x_A)
=\bigl|\Psi_{\rm lab}(x_1,\dots,x_A)\bigr|^2
=\frac{1}{8\pi^2}\int d\Omega\,\hat{R}(\Omega)\,\rho^{\rm wf}_{\rm intr}(x_1,\dots,x_A),
\end{equation}
where the double Euler integral is reduced, via the group property of $\hat{R}$ and a change of variables, to a single orientational average over the relative-rotation kernel
\begin{equation}
\label{eq:rho-q-intr}
\rho^{\rm wf}_{\rm intr}(x_1,\dots,x_A)
=\frac{1}{8\pi^2\mathcal{N}_A^2}\int d\Omega'\,
\Psi^*_{\rm intr}(x_1,\dots,x_A)\,
\hat{R}(\Omega')\,\Psi_{\rm intr}(x_1,\dots,x_A).
\end{equation}
Thus $\rho^{\rm wf}_{\rm intr}$ contains interference between intrinsic states of different orientations absent in $\rho^{\rm rg}_{\rm intr}$. It is a body-axis transition-kernel representative, not a uniquely defined intrinsic density of the projected state; its outer average in Eq.~\eqref{eq:rho-q-lab} is the laboratory density.

Because nucleons are fermions, $\Psi_{\rm intr}$ must be antisymmetric under the interchange of any two nucleons. 
In what follows we specialize to antisymmetrized product states for even-even symmetric nuclei (with equal numbers of protons and neutrons). We will show that the laboratory one- and two-body densities required by Eq.~\eqref{eq:glauber-phase} become explicit functionals of a finite set of single-particle orbitals. 

We construct $\Psi_{\rm intr}$ explicitly via a Slater determinant built from orthonormal single-particle orbitals \(\psi_j(x)\):
\begin{equation}
\label{eq:slater}
\Psi_{\rm intr}(x_1,\dots,x_A)
=\frac{1}{\sqrt{A!}}\det\bigl[\psi_j(x_i)\bigr]_{i,j=1}^A\;,
\end{equation}
where each single-particle orbital $\psi_j(x)$ can be factorized as a direct product of a spatial wave function and a spin-isospin wave function,
\begin{equation}
\label{eq:orb-factor}
\psi_j(x)=\phi_k(\mathbf{r})\otimes\chi_\alpha(s,\tau)\;. 
\end{equation}
Here the composite index $j$ runs over both the spatial orbital label $k=1,\dots,A/g$ and the internal label $\alpha=1,2,3,4$. Consequently each spatial orbital $\phi_k$ is occupied by four nucleons, giving the degeneracy factor $g=4$ when summing over internal degrees of freedom. All spatial one- and two-body densities below are understood after this spin-isospin trace.

In the rigid-rotor construction the reduced densities follow from the Slater algebra of the occupied spatial orbitals.
We introduce the spatial one-body density matrix~\cite{Lowdin:1955zzb,Lowdin:1955zza}
\begin{equation}
\label{eq:gamma}
\gamma(\mathbf{r},\mathbf{r}')
=\sum_{k=1}^{A/g}\phi_k(\mathbf{r})\phi_k^*(\mathbf{r}').
\end{equation}
Tracing the internal degrees of freedom then yields the rigid-rotor one-body density
\begin{equation}
\label{eq:rho1-intr-classical}
\rho^{\rm rg}_{1,\rm intr}(\mathbf{r})=g\,\gamma(\mathbf{r},\mathbf{r}),
\end{equation}
and the rigid-rotor two-body density
\begin{equation}
\label{eq:rho2-intr-classical}
\rho_{2,{\rm intr}}^{\rm rg}(\mathbf{r},\mathbf{r}')
=\rho_{1,\rm intr}^{\rm rg}(\mathbf{r})\,\rho_{1,\rm intr}^{\rm rg}(\mathbf{r}')
-g\bigl|\gamma(\mathbf{r},\mathbf{r}')\bigr|^2.
\end{equation}
Both terms already include the full sum over spin and isospin.
The first is the direct contribution: after that sum it is simply the product of the two one-body densities.
The second is the exchange contribution: it encodes the Pauli principle, and because only nucleons with the same spin-isospin labels block one another, it carries a single factor of $g$ rather than $g^2$.
Equation~\eqref{eq:rho2-intr-classical} is normalized to \(A(A-1)\).
At the same position $\mathbf{r}=\mathbf{r}'$, Pauli exclusion therefore suppresses the two-body density only partially,
\begin{equation}
\label{eq:rho2-coinc-classical}
\rho_{2,{\rm intr}}^{\rm rg}(\mathbf{r},\mathbf{r})
=\frac{g-1}{g}\,[\rho_{1,\rm intr}^{\rm rg}(\mathbf{r})]^2. 
\end{equation}
For $g=4$ that four internal states share each spatial orbital, the factor is 3/4: pairs of nucleons at same points are reduced by 1/4 but not forbidden. Equations~\eqref{eq:rho1-intr-classical}-\eqref{eq:rho2-coinc-classical} are intrinsic densities; the corresponding laboratory densities follow from the Haar average in Eq.~\eqref{eq:rho-c-lab}, with no additional weight.

For the wave-function-rotor construction the same Slater determinant is substituted into the relative-rotation kernel~\eqref{eq:rho-q-intr}.
Relative rotation renders the bra and ket orbital sets mutually non-orthogonal, so the reduced densities are those of non-orthogonal Slater determinants~\cite{Lowdin:1955zzb,Lowdin:1955zza}. 
The wave-function-rotor counterpart of the one-body density matrix $\gamma$ is the reduced one-body transition density matrix between $\Psi_{\rm intr}$ and its relatively rotated copy,
\begin{equation}
\label{eq:Gamma}
\Gamma(\Omega',\mathbf{r},\mathbf{r}')
= \sum_{k,l}^{A/g} \phi_k^*(\mathbf{r})\, \left[M^{-1}(\Omega')\right]_{lk}\, \hat{R}(\Omega')\, \phi_l(\mathbf{r}'),
\end{equation}
built from the orbital transfer matrix
\begin{equation}
\label{eq:M-matrix}
M_{kl}(\Omega') = \int d^3\mathbf{r}\, \phi_k^*(\mathbf{r})\, \hat{R}(\Omega')\, \phi_l(\mathbf{r}).
\end{equation}
When $\Omega'=0$ one has $M=\mathbf{I}$ and $\Gamma(0,\mathbf r,\mathbf r')=\gamma(\mathbf r',\mathbf r)$. Strictly speaking, a rotation of the many-body state should act jointly on spatial and spin degrees of freedom. In the spin-isospin-saturated occupancy used here, the occupied spin space is invariant under rotation, so $\Gamma$ and $M$ can be evaluated with spatial rotation alone.
After the spin-isospin sum, the body-axis transition kernels become weighted averages over relative orientations,
\begin{equation}
\label{eq:rho1-intr-quantum}
\rho^{\rm wf}_{1,\rm intr}(\mathbf{r})
= g\, \frac{\int d\Omega'\, [\det M(\Omega')]^g \, \Gamma(\Omega',\mathbf{r},\mathbf{r})}
{\int d\Omega'\, [\det M(\Omega')]^g},
\end{equation}
\begin{equation}
\label{eq:rho2-intr-quantum}
\rho^{\rm wf}_{2,{\rm intr}}(\mathbf{r},\mathbf{r}')
= \frac{\int d\Omega'\, [\det M(\Omega')]^g\,
\Bigl( g^2 \Gamma(\Omega',\mathbf{r},\mathbf{r})\,
\Gamma(\Omega',\mathbf{r}',\mathbf{r}')
- g\,\Gamma(\Omega',\mathbf{r},\mathbf{r}')\,
\Gamma(\Omega',\mathbf{r}',\mathbf{r}) \Bigr)}
{\int d\Omega'\, [\det M(\Omega')]^g}.
\end{equation}
Inside the integrand of Eq.~\eqref{eq:rho2-intr-quantum}, the exchange contraction term $\Gamma(\Omega',\mathbf r,\mathbf r')\Gamma(\Omega',\mathbf r',\mathbf r)$ appears because a transition density is generally non-Hermitian at nonzero relative rotation. This form gives the required $A(A-1)$ pair normalization.
The weight $[\det M]^g$ likewise folds the fourfold internal degeneracy into the orbital-transfer determinant, and is the concrete realization of the renormalization encoded by $\mathcal{N}_A$. 
Laboratory densities are again recovered by the orientational average in Eq.~\eqref{eq:rho-q-lab}.

%====================================================================
\section{Ne-20 in Nilsson model: densities and deformation}
\label{sec:nilsson}

We now take an example for the rigid-rotor and wave-function-rotor constructions of Sec.~\ref{sec:formal}: the light, strongly prolate nucleus Ne-20 described by an axially deformed harmonic-oscillator (Nilsson-type) potential.
The model is simple enough for closed-form one-body densities and an explicit orbital transfer matrix, yet rich enough that deformation, Pauli blocking effects, and relative-rotation weights can be varied or studied systematically.

The single-particle potential is
\begin{equation}
\label{eq:Nilsson-V}
V(x,y,z)=\frac12 m\omega_\perp^2(x^2+y^2)+\frac12 m\omega_z^2 z^2,
\end{equation}
with Cartesian oscillator energies
\begin{equation}
\label{eq:Nilsson-E}
E_{n_x n_y n_z}
=\hbar\omega_\perp(n_x+n_y+1)
+\hbar\omega_z\Bigl(n_z+\frac12\Bigr).
\end{equation}
Deformation is parametrized by the frequency ratio
\begin{equation}
\label{eq:kappa-def}
\kappa=\frac{\omega_z}{\omega_\perp}
=\Bigl(\frac{\alpha_z}{\alpha_\perp}\Bigr)^2,
\qquad
\alpha_\perp=\sqrt{\frac{m\omega_\perp}{\hbar}},
\qquad
\alpha_z=\sqrt{\frac{m\omega_z}{\hbar}}.
\end{equation}
Prolate shapes correspond to $\omega_z<\omega_\perp$, i.e.\ $\kappa<1$: the $z$ direction is softer, so the density elongates along the body axis.
We take $1/\alpha_\perp$ as the unit of length, so that all coordinates and densities hereafter are dimensionless.
When physical units are required we use the conversion scale $1/\alpha_\perp$=1.85 fm , which places the Ne-20 radius near 3 fm.

The ground-state filling of Ne-20 depends on $\kappa$. 
To keep the occupation ordered as $E_{000}<E_{001}<E_{100}=E_{010}<E_{002}$, we restrict $1/3<\kappa<1$. 
With spin-isospin degeneracy $g=4$, the five spatial orbitals in this window accommodate exactly 20 nucleons in Ne-20 and define our model space.

Rotations are conveniently handled in eigenstates of $\hat{L}_z$ with eigenvalues $m_z\hbar$. 
The five occupied orbitals, written through the Gaussian envelope $\phi_1$, are
\begin{align}
\label{eq:psi-five}
\phi_1
&=|000\rangle
=\Bigl(\frac{\kappa}{\pi^3}\Bigr)^{1/4}
\exp\Bigl[-\tfrac12(x^2+y^2+\kappa z^2)\Bigr],
\nonumber\\
\phi_2
&=|001\rangle
=\sqrt{2\kappa}\,z\,\phi_1,
\nonumber\\
\phi_3
&=\frac1{\sqrt2}\bigl(|100\rangle+i|010\rangle\bigr)
=(x+iy)\phi_1,
\nonumber\\
\phi_4
&=\frac1{\sqrt2}\bigl(|100\rangle-i|010\rangle\bigr)
=(x-iy)\phi_1,
\nonumber\\
\phi_5
&=|002\rangle
=\Bigl(\sqrt{2}\,\kappa z^2-\frac1{\sqrt2}\Bigr)\phi_1.
\end{align}
Here $\phi_3$ and $\phi_4$ have $m_z=\pm 1$ respectively and are combinations of the first excited in-plane states, $\phi_2$ and $\phi_5$ carry the axial excitation responsible for the prolate deformation.

In the rigid-rotor picture one fills these five orbitals and forms $\rho^{\rm rg}_{1,\rm intr}=g\gamma(\mathbf{r},\mathbf{r})$  according to Eq.~\eqref{eq:rho1-intr-classical} as 
\begin{equation}
\label{eq:rho1-Ne-classical}
\rho^{\rm rg}_{1,\rm intr}(\mathbf{r})
=4\Bigl(\frac{\kappa}{\pi^3}\Bigr)^{1/2}
\exp\bigl[-(x^2+y^2+\kappa z^2)\bigr]
\Bigl[\tfrac32+2(x^2+y^2)+2\kappa^2 z^4\Bigr],
\end{equation}
which normalized to normalized to 20 (the degeneracy $g=4$ is included). We see in Eq. (\ref{eq:rho1-Ne-classical}) that the anisotropic Gaussian form sets the overall shape, and the polynomial in brackets is the contribution from all occupied orbitals.

\begin{figure}[htbp]
\centering
\includegraphics[width=0.6\textwidth]{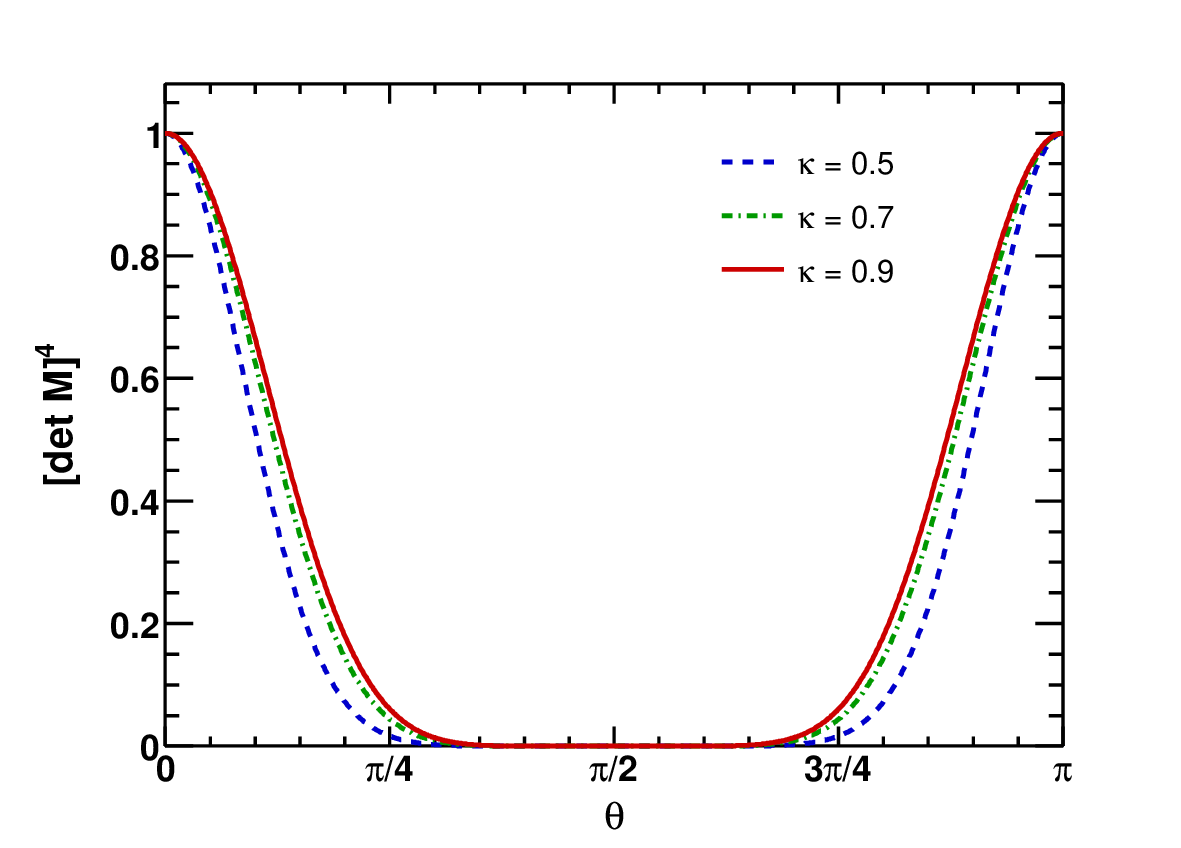}
\caption{Dependence of $[\det M(\theta,\kappa)]^4$ on $\theta$ for three values of $\kappa$. 
The curves for $\kappa=0.5$, 0.7, and 0.9 are shown as the blue dashed, green dash-dotted, and red solid lines, respectively.
In the limit $\kappa\to 1$, $[\det M(\theta,\kappa)]^4\to\cos^8\theta$, which nearly coincides with the red (\(\kappa=0.9\)) curve.}
\label{fig:detM4-vs-theta}
\end{figure}

For the wave-function-rotor construction the same orbitals enter the relative-rotation kernel through the orbital transfer matrix $M_{kl}(\Omega')$ in Eq.~\eqref{eq:M-matrix}.
In evaluating $M_{kl}(\Omega)$ (in the following discussion we use $\Omega$ in place of $\Omega'$ just for notational convenience), we note that the operator $\hat{R}(\Omega)=\hat{R}_z(\phi)\hat{R}_y(\theta)\hat{R}_z(\alpha)$ acts on $\phi_l(\mathbf{r})$ and the occupied orbitals are eigenstates of $\hat{L}_z$.
The effect of $\hat{R}_z(\alpha)$ on $\phi_l(\mathbf{r})$ is to produce a phase $e^{-im_z\alpha}$.
Because the five occupied orbitals have total $m_z=0$, these phases multiply to unity in the ket Slater determinant. Thus $\hat{R}_z(\alpha)$ leaves that many-body state unchanged.
% although it cannot be dropped from each individual $M_{kl}(\Omega)$. 
 The remaining $\hat{R}_z(\phi)$ dependence is included by an azimuthal average when the body-axis transition kernels are evaluated.
The nontrivial relative tilt is therefore represented by $\hat{R}_y(\theta)=e^{-i\theta\hat{L}_y}$.
In the five-orbital basis this defines the orbital transfer matrix
\begin{equation}
\label{eq:M-theta}
M_{ij}(\theta,\kappa)
=\int d^3\mathbf{r}\,
\phi_i^*(\mathbf{r})\,e^{-i\theta\hat{L}_y}\,\phi_j(\mathbf{r}).
\end{equation}
So $M$ is a real $5\times5$ matrix in the basis~\eqref{eq:psi-five} with the determinant in analytical form 
\begin{equation}
\label{eq:detM}
\det M(\theta,\kappa)
=65536\sqrt{2}\,\cos^2\theta
\Bigl [\frac{\kappa}{1+6\kappa+\kappa^2-(\kappa-1)^2\cos 2\theta}\Bigr]^{11/2}.
\end{equation}
As discussed after Eq.~\eqref{eq:rho1-intr-quantum}, the transition kernels in the wave-function-rotor picture involve the weight $[\det M]^g$. For $A=20$ and $g=4$ this is $[\det M(\theta,\kappa)]^4$, whose angular integral fixes $\mathcal{N}_A$ in the present model.
In the spherical limit $\kappa\to 1$ one has $[\det M]^4\to\cos^8\theta$. For prolate $\kappa$ the weight remains peaked near aligned orientations and is suppressed toward side-on relative rotations ($\theta\to\pi/2$), where $\det M\to 0$. Figure~\ref{fig:detM4-vs-theta} shows this angular weight for three representative deformations.

The matrix $M(\theta,\kappa)$ encodes the relative tilt, but does not by itself define the projected transition kernel.
The $\hat{L}_y$ rotation mixes $m_z$ components, so at fixed $\theta$ the spatial densities still depend on the azimuthal orientation of the tilted body axis. 
The remaining rotation about $z$ is implemented as an average over $\phi$.
The transition kernels are built at fixed $\theta$ from $M(\theta ,\kappa )$ and $M^{-1}(\theta ,\kappa )$ and by average over $\theta$ with the weight $[\det M(\theta ,\kappa )]^4$ and over $\phi$. Finally the full laboratory Haar average of Sec.~\ref{sec:formal} is taken.

\begin{figure}[htbp]
\centering
\begin{minipage}[t]{0.32\textwidth}
\centering
\includegraphics[width=\textwidth]{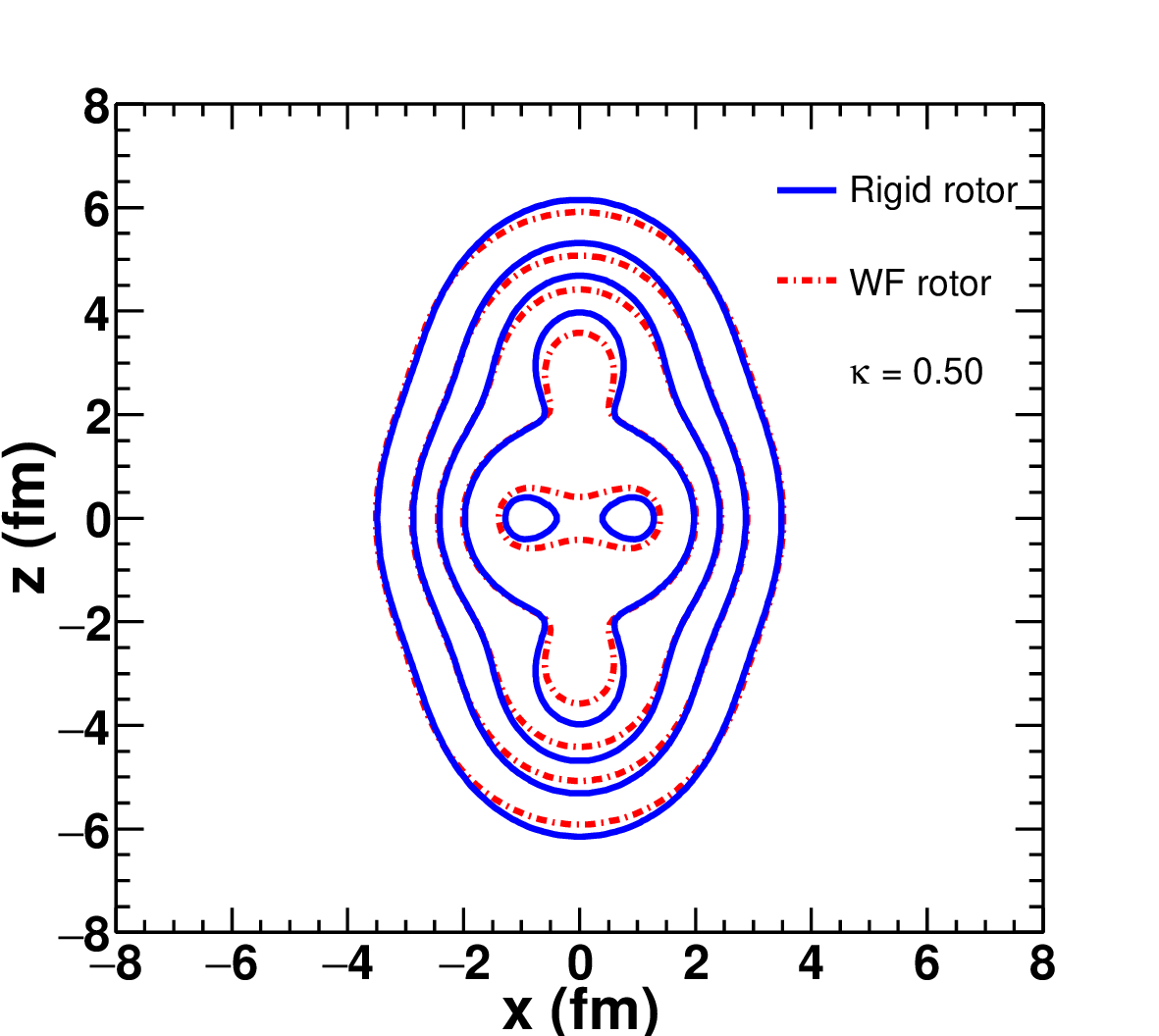}
\end{minipage}\hfill
\begin{minipage}[t]{0.32\textwidth}
\centering
\includegraphics[width=\textwidth]{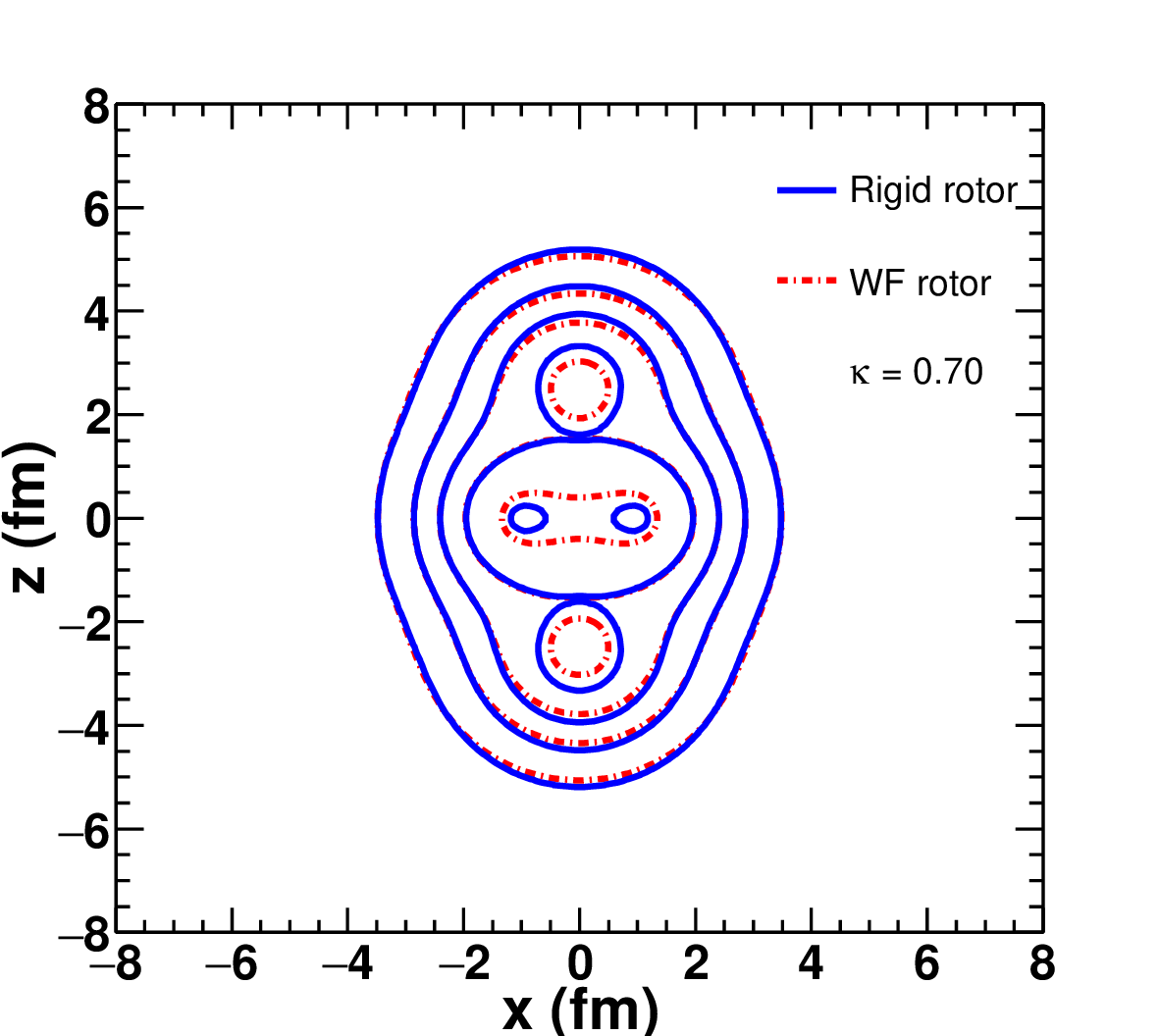}
\end{minipage}\hfill
\begin{minipage}[t]{0.32\textwidth}
\centering
\includegraphics[width=\textwidth]{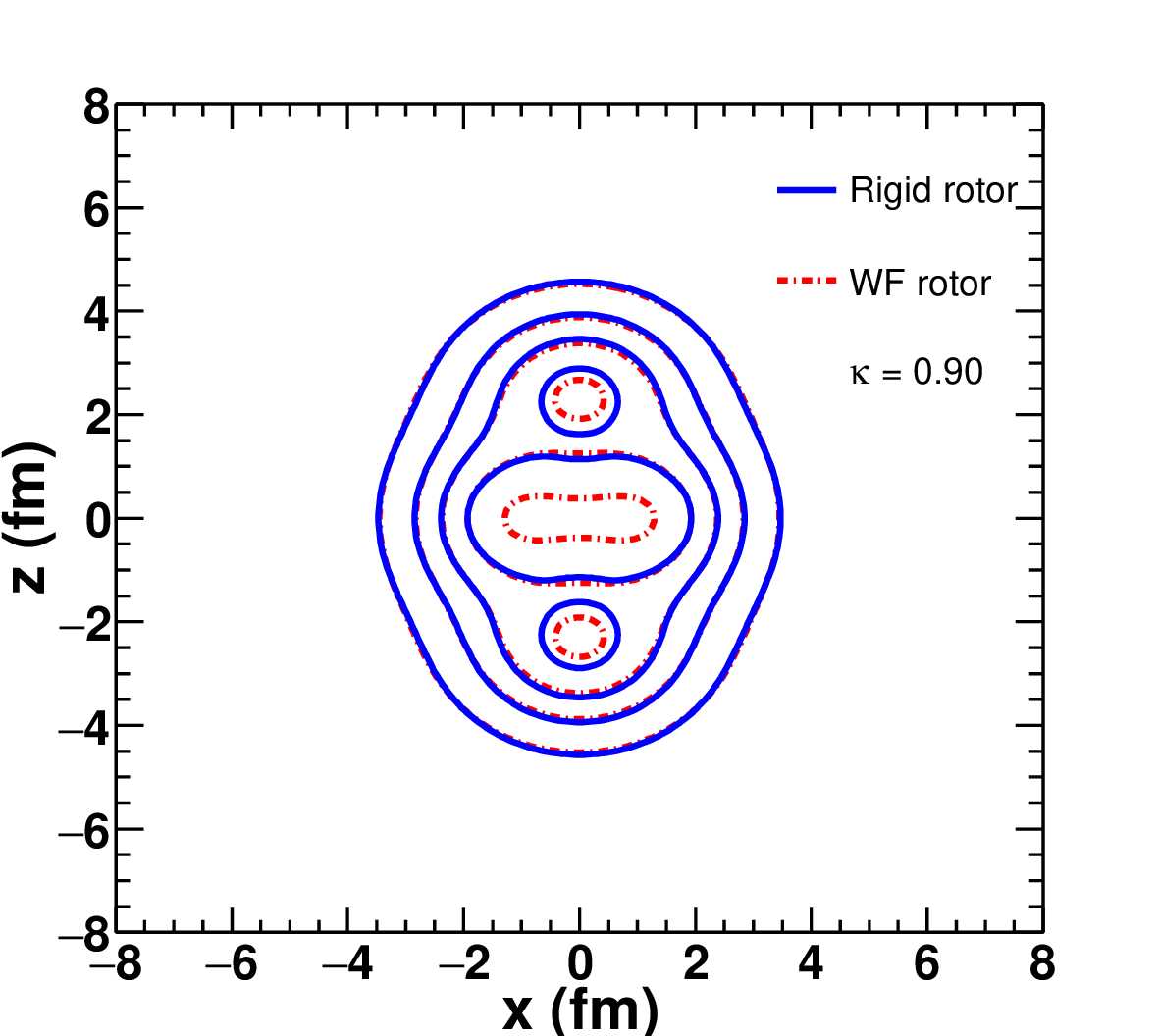}
\end{minipage}
\caption{Body-axis one-body contours for $\kappa=0.5$, 0.7, and 0.9 (left to right). Blue solid lines show the rigid-rotor intrinsic density; red dashed lines show the wave-function-rotor transition-kernel representative.}
\label{fig:intr-rho1-overlay}
\end{figure}

The rigid-rotor intrinsic density~\eqref{eq:rho1-Ne-classical} can be compared with the projected body-axis kernel~\eqref{eq:rho1-intr-quantum}.
Figure~\ref{fig:intr-rho1-overlay} shows the corresponding contours in the intrinsic $xz$ plane at three values $\kappa=0.5$, 0.7, and 0.9.
Blue solid and red dashed curves are the rigid-rotor and wave-function-rotor results, respectively.
In each panel the levels are drawn at 95\%, 75\%, 55\%, 35\%, and 15\% of a common peak value, taken as the maximum of the wave-function-rotor representative so that the two sets of contours are directly comparable.

The two body-axis functions are qualitatively similar, and as $\kappa$ decreases the prolate elongation of both sets of contours grows. The outer contours are approximately prolate ellipsoidal, while the interior develops an equatorial ring, visible in the $xz$ plane as a waist. Compared with the rigid-rotor result at the same $\kappa$, the wave-function-rotor outer contours contract along the symmetry axis and the central value rises; the equatorial ring is attenuated and the inner contours become smoother. %These contour differences do not change the starting Nilsson orbitals. 

Orientational averaging then yields the laboratory-frame one-body densities.
Consistent with the intrinsic-frame trends above, the wave-function-rotor laboratory density is enhanced near the center and depleted in the periphery relative to the rigid-rotor result. For $\kappa$ close to unity, however, the radial spreading of the equatorial ring reverses this ordering at large radii, where the wave-function-rotor density again exceeds the rigid-rotor one.

\begin{figure}[htbp]
\centering
\begin{minipage}[t]{0.48\textwidth}
\centering
\includegraphics[width=\textwidth]{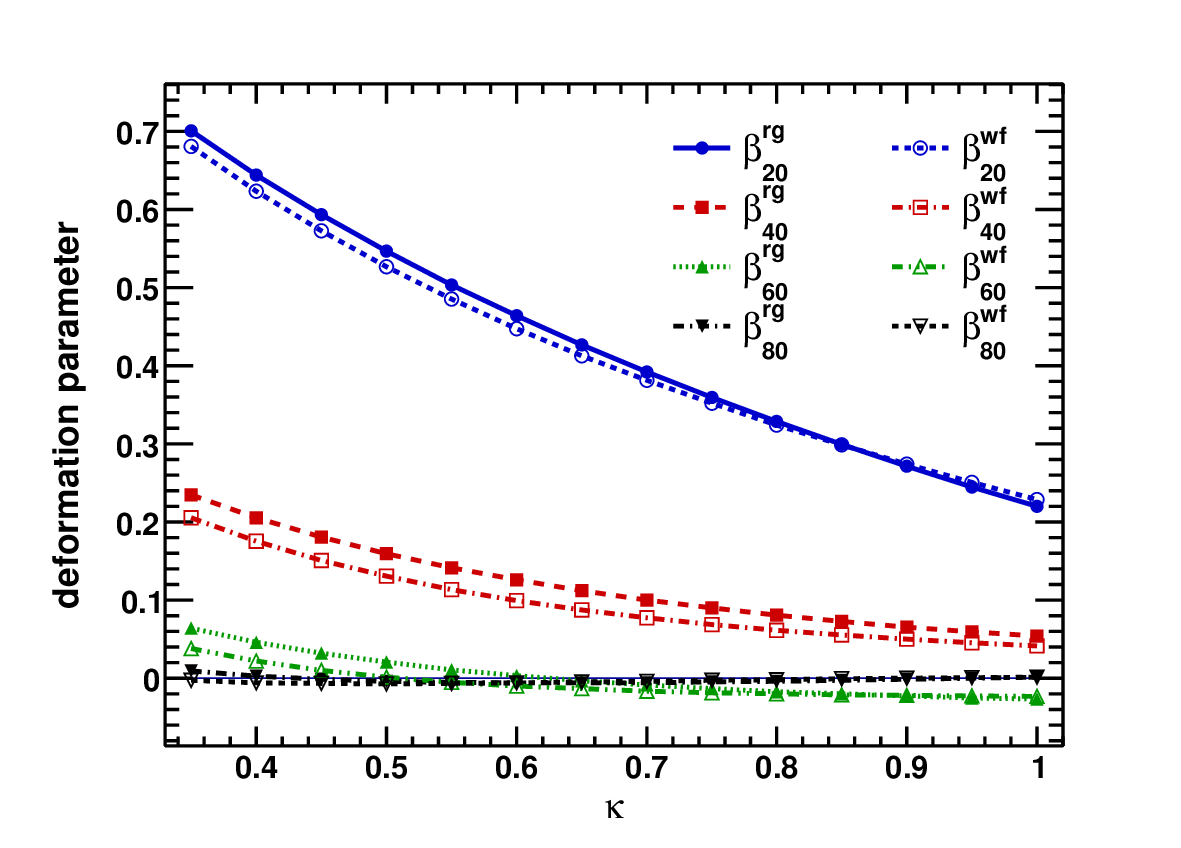}
\end{minipage}\hfill
\begin{minipage}[t]{0.48\textwidth}
\centering
\includegraphics[width=\textwidth]{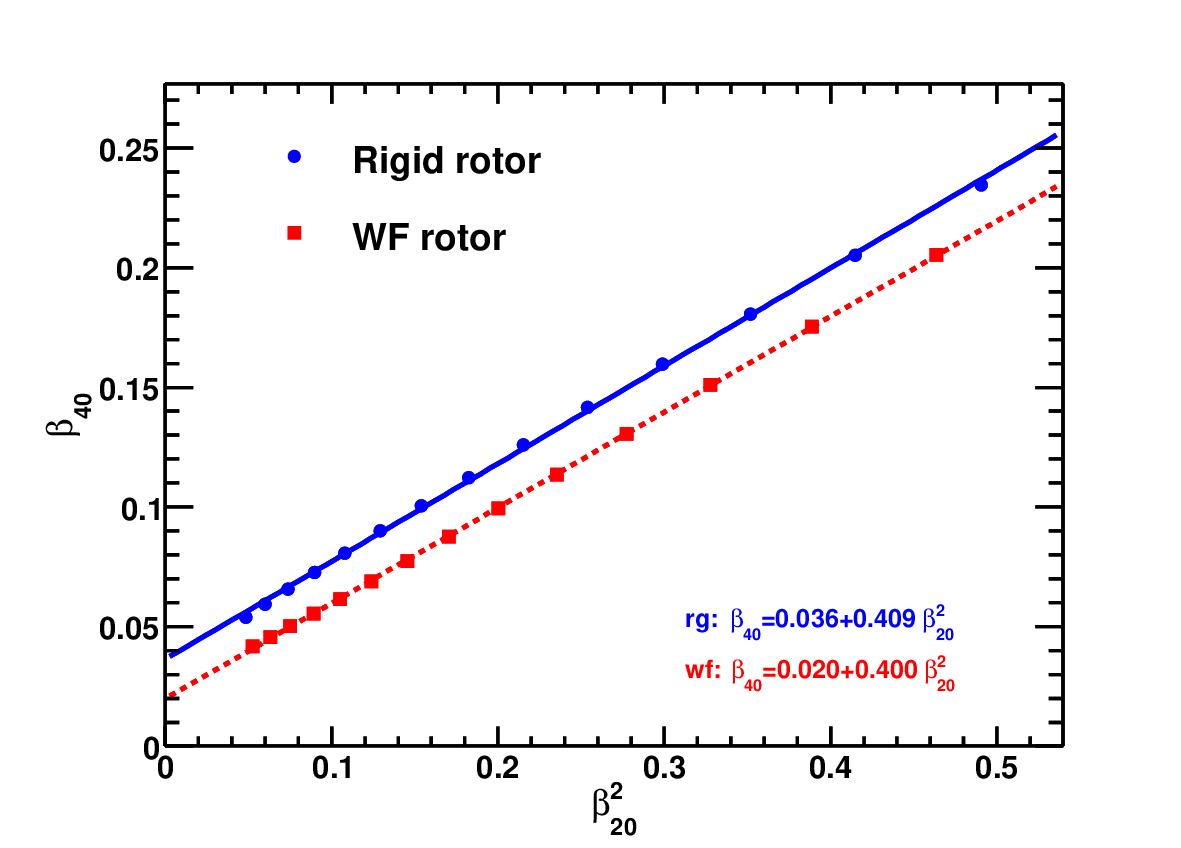}
\end{minipage}\hfill
\caption{Body-axis contour multipole coefficients of Ne-20.
Left: $\beta_{\ell 0}$ for $\ell=2,4,6,8$ versus $\kappa$ in the rigid-rotor and wave-function-rotor cases.
Right: $\beta_{40}$ versus $\beta_{20}^2$ for same parameters, with linear fits to Eq.~\eqref{eq:beta40-vs-beta20sq}.}
\label{fig:beta-class-quan-fit}
\end{figure}

From the displayed body-axis functions one can also extract contour multipoles. 
An outer isodensity contour $R(\theta)$ is defined by 
\begin{equation}
\label{surface}
\rho_{1,\rm intr}\bigl(R(\theta),\theta;\kappa\bigr)=f\,\rho_{\mathrm{peak}}\;,
\end{equation}
where $\rho_{\mathrm{peak}}$ is a common peak value used for the contour levels in Fig.~\ref{fig:intr-rho1-overlay} and chosen to be the maximum of the wave-function-rotor representative, and $f$ is the fraction of this value for the outermost contour in Fig.~\ref{fig:intr-rho1-overlay}. As we can see later that choosing a different value of $f$ will not bring significant difference to the main results as long as $f\ll 1$, i.e. it corresponds to the region near the edge of the contour. Expanding this contour in axial spherical harmonics gives 
\begin{equation}
\label{eq:R-Ylm-expand}
R(\theta) = R_0\Bigl[
1+\beta_{20} Y_{20}
+\beta_{40} Y_{40}
+\beta_{60} Y_{60}
+\beta_{80} Y_{80}
\Bigr]\;,
\end{equation}
where $\beta_{n0}$ are contour coefficients. For the wave-function rotor they characterize the chosen transition-kernel representative, rather than an intrinsic density observable of the projected state. The left panel of Fig.~\ref{fig:beta-class-quan-fit} shows these coefficients versus $\kappa$ in the rigid-rotor and wave-function-rotor cases.

We see in the left panel of Fig.~\ref{fig:beta-class-quan-fit} that as $\kappa$ increases, the contour coefficients decrease in magnitude. The quadrupole coefficient dominates over higher multipoles. 
Even at $\kappa=1$, a residual $\beta_{20}$ remains: the oscillator potential is then spherical, but the occupied orbitals still generate a nonspherical density through their polynomial structure. 
The rigid-rotor contour coefficients generally lie above the wave-function-rotor ones, and for $\beta_{40}$ this gap is more pronounced. In this parameter space the two leading coefficients are correlated. The hexadecapole coefficient taken from an outer contour is related to the quadrupole coefficient by
\begin{equation}
\label{eq:beta40-vs-beta20sq}
\beta_{40}=p+q\beta_{20}^{2},
\end{equation}
to a very good precision, the residuals lie at the level of $10^{-3}$, well below the magnitude of $\beta_{40}$. 
A linear fit gives $q\simeq 0.4$ for both the rigid-rotor and wave-function-rotor surfaces, with little sensitivity to the outer-contour fraction $f$. The intercept $p$ in the rigid-rotor case is larger than that in the wave-function-rotor case by about $0.015$, which accounts for the difference between the two constructions.

%====================================================================
\section{Eccentricity in ultracentral collisions: two-body correlations}
\label{sec:ecc}

In high-energy nuclear collisions, the final-state anisotropic flow $v_n$ can be measured experimentally, while the initial-state geometric eccentricity $\varepsilon_n$ can be computed from the nuclear densities that set the collision geometry. We use the leading-order ultracentral mapping of Ref.~\cite{Duguet:2025hwi} to compare the two constructions through $\langle\varepsilon_n^2\rangle$. Relating this quantity to measured $\langle v_n^2\rangle$ requires a collision and medium-response calculation.

In central nuclear collisions, an incoming nucleus is characterized in the transverse plane by the thickness function $t(\mathbf{x})=\sum_{i=1}^{A}h(\mathbf{x}-\mathbf{r}_{\perp i})$, which stacks the three-dimensional nucleon density onto the transverse plane~\cite{Duguet:2025hwi}. Here $\mathbf{r}_{\perp i}$ is the transverse coordinate the $i$-th nucleon in the nucleus, and $h$ is the two-dimensional nucleon profile seen by a high-energy probe. Nucleon coordinates are sampled from the laboratory many-body distribution of the construction being studied, so that event-by-event fluctuations of $t(\mathbf{x})$ encode its many-body correlations. 
The initial entropy density $s(\mathbf{x})$ of the fireball in collisions is assumed to depend locally on the thickness functions $t(\mathbf{x})$ and $t'(\mathbf{x})$ of two colliding nuclei. For identical nuclei in ultracentral collisions it is assumed that $s(\mathbf{x})$ is proportional to  $\bigl[t(\mathbf{x})+t'(\mathbf{x})\bigr]/2$. 
Under this mapping, the one- and two-point correlators of the entropy density, $C_1(\mathbf{x})=\langle s(\mathbf{x})\rangle$ and $C_2(\mathbf{x}_1,\mathbf{x}_2)=\langle s(\mathbf{x}_1)s(\mathbf{x}_2)\rangle-C_1(\mathbf{x}_1)C_1(\mathbf{x}_2)$, are given by convolutions of the one- and two-body densities $\rho_{1,\rm lab}(\mathbf{r})$ and $\rho_{2,\rm lab}(\mathbf{r}_1,\mathbf{r}_2)$ with the nucleon profile $h$.

The coordinate-space eccentricity $\mathcal{E}_n$ for a given entropy profile $s(\mathbf{x})$ is defined as 
\begin{equation}
\mathcal{E}_n
=-\frac{\displaystyle\int|\mathbf{x}|^n\,e^{in\phi}\,s(\mathbf{x})\,d^2x}
{\displaystyle\int|\mathbf{x}|^n\,s(\mathbf{x})\,d^2x},
\qquad
\varepsilon_n=|\mathcal{E}_n|\;,
\end{equation}
where $\phi$ is the angle of $\mathbf{x}$. Expanding about the mean entropy profile to leading nontrivial order in $\delta s(\mathbf{x})=s(\mathbf{x})-\langle s(\mathbf{x})\rangle$ then yields~\cite{Duguet:2025hwi}
\begin{equation}
\label{eq:eps-C2}
\langle\varepsilon_n^2\rangle
\simeq\frac{\displaystyle\int|\mathbf{x}_1|^n\,|\mathbf{x}_2|^n\,e^{in(\phi_1-\phi_2)}\,C_2(\mathbf{x}_1,\mathbf{x}_2)\,d^2x_1\,d^2x_2}
{\displaystyle\Bigl(\int|\mathbf{x}|^n\,C_1(\mathbf{x})\,d^2x\Bigr)^2}.
\end{equation}
We see that the mean squared anisotropy is determined by the connected correlator $C_2(\mathbf{x}_1,\mathbf{x}_2)$ in the numerator and by $C_1(\mathbf{x})$ in the denominator. 
In the pointlike limit $h(\mathbf{x})\to\delta^{(2)}(\mathbf{x})$ that nucleon size is negligible in the nucleus and with the normalization conditions 
\begin{align}
\int\rho_{1,\rm lab}(\mathbf{r})\,d^3r=&A \;,\nonumber\\
\iint\rho_{2,\rm lab}(\mathbf{r}_1,\mathbf{r}_2)\,d^3r_1d^3r_2=&A(A-1)\;,
\end{align}
Eq.~\eqref{eq:eps-C2} is reduced to the form 
\begin{equation}
\label{eq:eps-decomp}
\langle\varepsilon_n^2\rangle
\simeq\frac{\langle \hat{R}_{2n} \rangle+\langle \hat{\mathcal{E}}_{n} \rangle}{2\,\langle \hat{R}_{n} \rangle^{\,2}},
\end{equation}
where $\langle \hat{R}_{n} \rangle$, $\langle \hat{R}_{2n} \rangle$ and $\langle \hat{\mathcal{E}}_{n} \rangle$ are evaluated from the corresponding laboratory one- and two-body densities:
\begin{align}
\langle \hat{R}_{n} \rangle
&=\int\rho_{1,{\rm lab}}(\mathbf{r})\,r_\perp^{\,n}\,d^3r,
\qquad
\langle \hat{R}_{2n} \rangle=\int\rho_{1,{\rm lab}}(\mathbf{r})\,r_\perp^{\,2n}\,d^3r, \nonumber\\
\langle \hat{\mathcal{E}}_{n} \rangle
&=\int\rho_{2,{\rm lab}}(\mathbf{r}_1,\mathbf{r}_2)\,
r_{1\perp}^n r_{2\perp}^n\,e^{in(\phi_1-\phi_2)}\,d^3r_1d^3r_2 \;.
\end{align}
Here $r_{1\perp}^n r_{2\perp}^n\,e^{in(\phi_1-\phi_2)}$ measures the $n$-th order azimuthal correlation of two nucleons in the transverse plane. The moments $\langle \hat{R}_{n} \rangle$ and $\langle \hat{R}_{2n} \rangle$ depend only on the one-body density $\rho_{1,\rm lab}(\mathbf{r})$. In the independent-source limit at fixed $A$, $\rho_{2,{\rm lab}}(\mathbf{r}_1,\mathbf{r}_2)=\rho_{1,{\rm lab}}(\mathbf{r}_1)\rho_{1,{\rm lab}}(\mathbf{r}_2)$, and $\langle \hat{\mathcal{E}}_{n} \rangle$ vanishes by the spherical symmetry of the laboratory one-body density, leaving the fluctuation term that scales as $\langle\varepsilon_n^2\rangle_{\rm indep} \simeq \langle \hat{R}_{2n} \rangle/(2\,\langle \hat{R}_{n} \rangle^{\,2}) \sim 1/A$. At finite impact parameter the reaction-plane eccentricity rapidly overwhelms this fluctuation-driven term, but it is suppressed in ultracentral collisions. Therefore, $\langle\varepsilon_n^2\rangle$ in this ultracentral approximation is sensitive to the harmonic components of the laboratory two-body density; relating it to measured $\langle v_n^2\rangle$ requires a collision and medium-response calculation~\cite{Duguet:2025hwi}.

\begin{figure}[htbp]
\centering
\begin{minipage}[t]{0.48\textwidth}
\centering
\includegraphics[width=\textwidth]{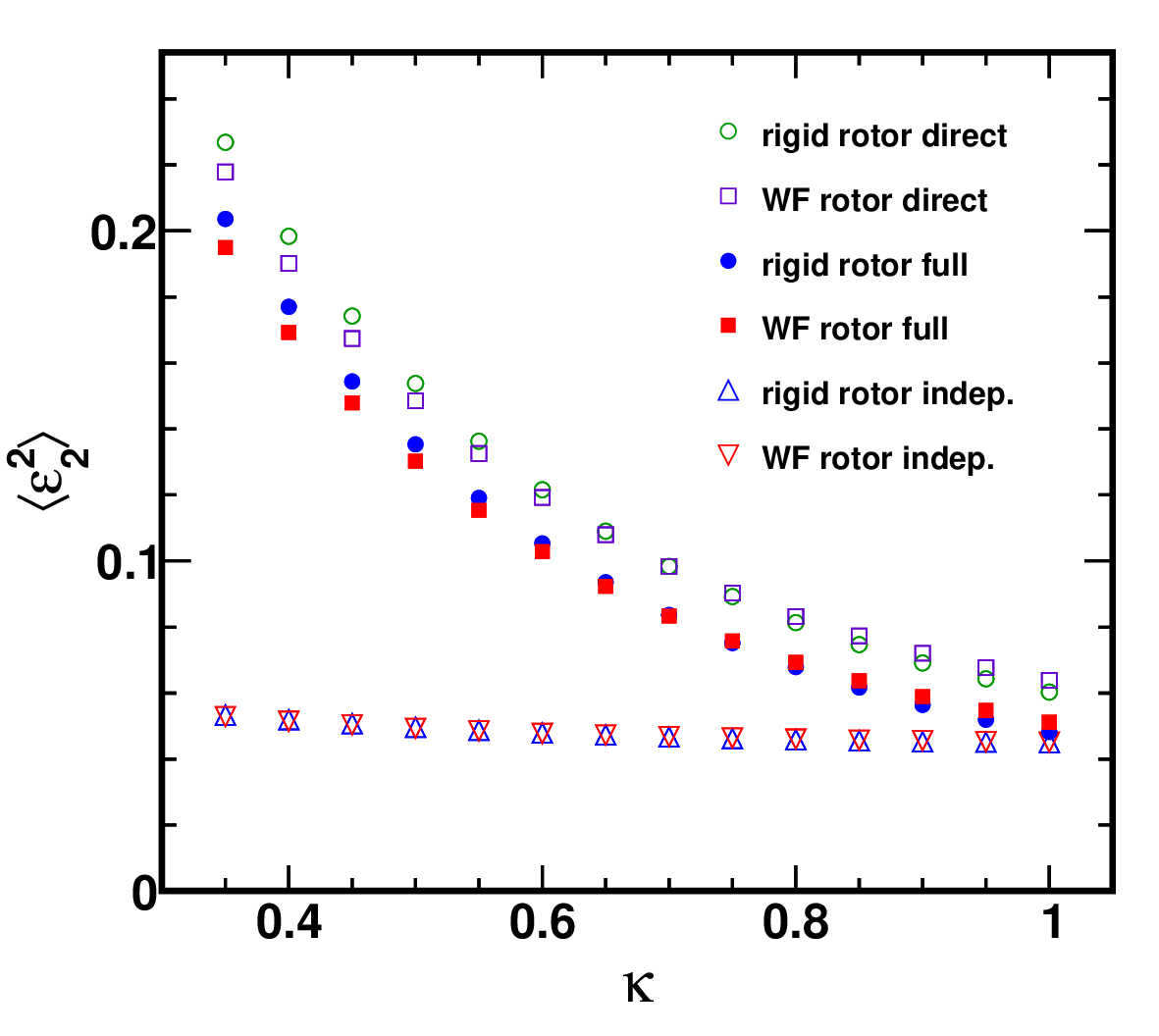}
\end{minipage}\hfill
\begin{minipage}[t]{0.48\textwidth}
\centering
\includegraphics[width=\textwidth]{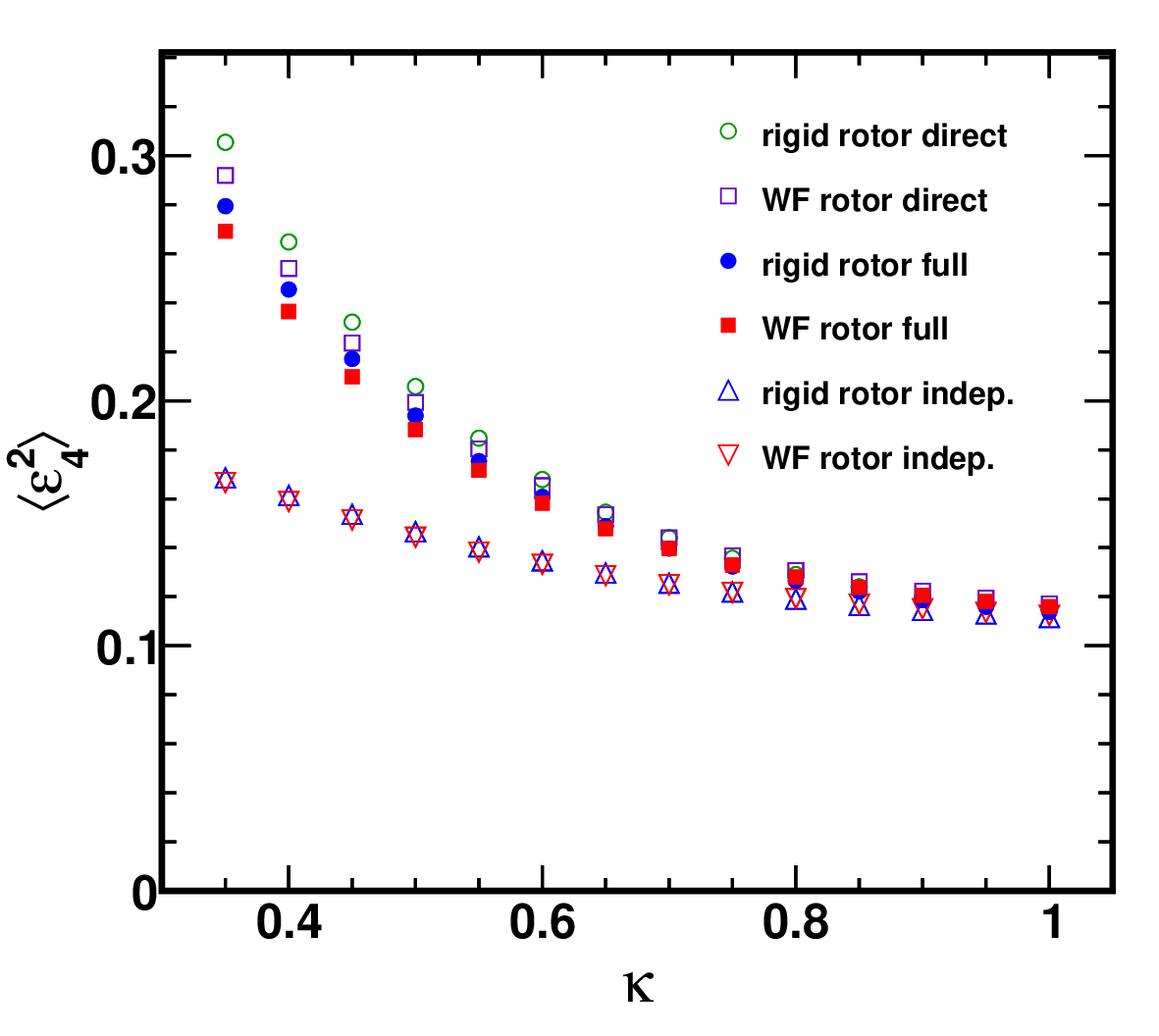}
\end{minipage}
\caption{Geometric eccentricity variance $\langle\varepsilon_n^2\rangle$ versus the deformation parameter $\kappa$ for Ne-20. The left and right panels show the quadrupole ($n=2$) and hexadecapole ($n=4$) variance, respectively. Open green circles and magenta squares show the rigid-rotor and wave-function-rotor results with only the direct contribution (without the exchange contribution). Filled blue circles and red squares show the rigid-rotor and wave-function-rotor results including both the direct and exchange contributions. Open blue and red triangles denote the corresponding independent-source baselines $\langle\varepsilon_n^2\rangle_{\mathrm{indep}}$.}
\label{fig:eps-vs-kappa}
\end{figure}

We evaluate $\langle\varepsilon_n^2\rangle$ through one- and two-body densities in laboratory frame for both rigid rotor and wave-function rotor cases. Figure \ref{fig:eps-vs-kappa} shows $\langle\varepsilon_n^2\rangle$ as a function of $\kappa$ for $n=2$ (left) and $n=4$ (right). In both cases the displayed $\langle\varepsilon_n^2\rangle$ decreases as $\kappa$ increases. The direct-only and full results vary strongly with $\kappa$, and exchange lowers the curves relative to the direct-only points.

For $n=2$ the independent-source baseline is essentially a constant of order $1/A$, as expected when connected two-body correlations are absent. For $n=4$ the baseline itself falls with $\kappa$. This does not mean that azimuthal correlations survive in the independent-source case, it simply reflects that the one-body transverse moments depend on the deformation parameter, and so does the ratio $\langle \hat{R}_{2n} \rangle /(2\,\langle \hat{R}_{n} \rangle ^{\,2})$.

Over the plotted $\kappa$ range we fit the calculated variance versus the contour coefficient by
\begin{equation}
\label{eq:eps-linear}
\langle\varepsilon_n^2\rangle =a_n+b_n\beta_{n0}^2,
\end{equation}
where $a_n$ and $b_n$ are fit parameters; the independent-source baseline is the one-body ratio $\langle \hat{R}_{2n} \rangle /(2\,\langle \hat{R}_{n} \rangle ^{\,2})$ and can vary across the parameter space.
A pure quadrupole Gaussian surface yields the benchmark result 
$\langle\varepsilon_2^2\rangle=1/A+(3/4\pi)\beta_{2}^2$ 
where $\beta_2$ is a different deformation parameter defined in Ref. \cite{Duguet:2025hwi} 
from $\beta_{20}$ in Eq.(\ref{surface}). 
The Gaussian benchmark does not include a separate $\beta_{40}$ parameter.

\begin{figure}[htbp]
\centering
\includegraphics[width=0.6\textwidth]{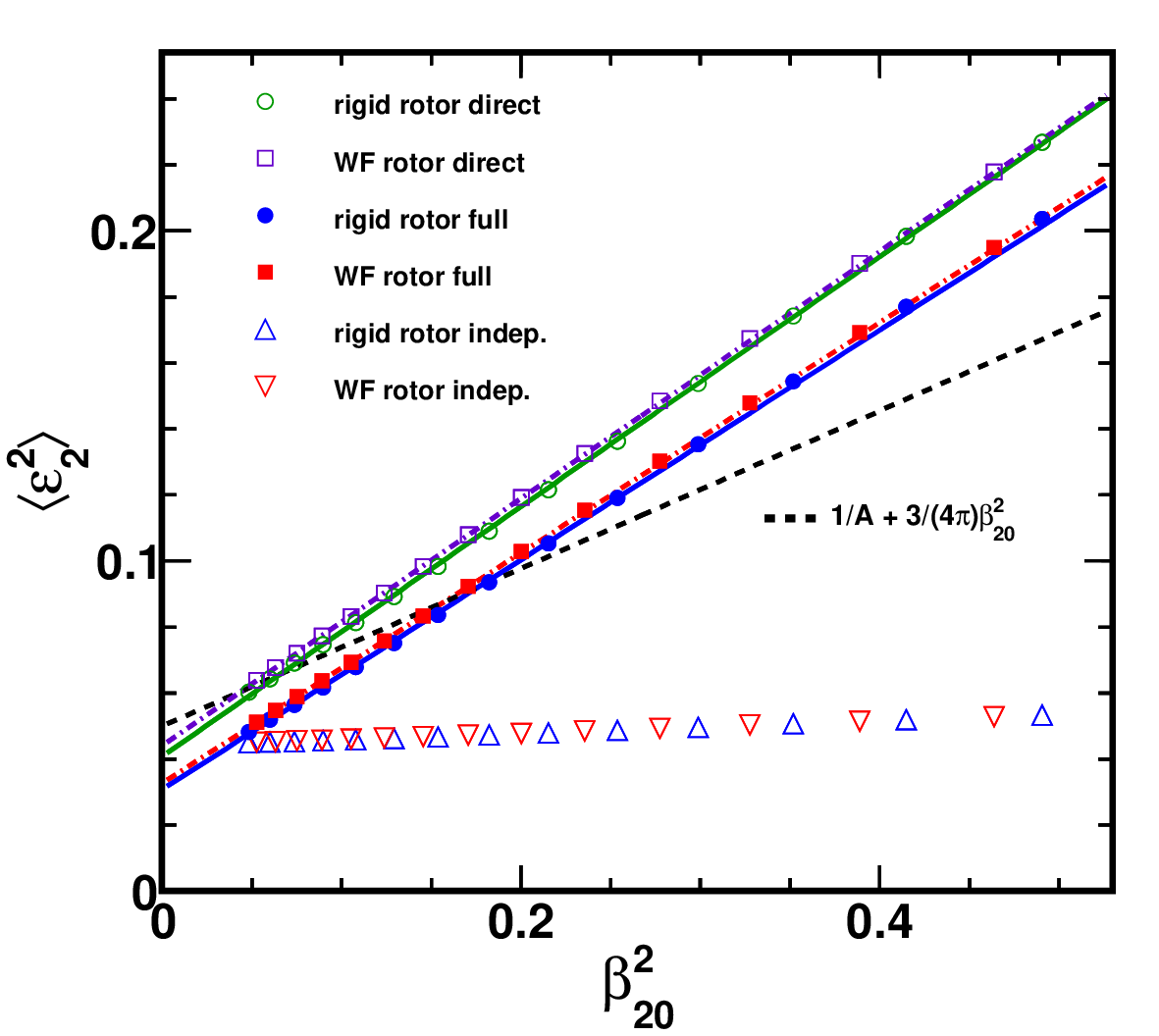}
\caption{Geometric eccentricity variance $\langle\varepsilon_2^2\rangle$ versus $\beta_{20}^2$ for Ne-20.
The symbols are the same as in Fig.~\ref{fig:eps-vs-kappa}.
The black dashed line is the Gaussian rigid-rotor result of Ref.~\cite{Duguet:2025hwi}, which does not include the exchange contribution. Solid and dash-dotted curves are fits to Eq.~\eqref{eq:eps-linear} for the rigid-rotor and wave-function-rotor results, respectively.}
\label{fig:eps2-vs-beta2surf}
\end{figure}

Figure \ref{fig:eps2-vs-beta2surf} presents the geometric eccentricity variance $\langle\varepsilon_2^2\rangle$ versus $\beta_{20}^2$. The rigid-rotor result of Ref.~\cite{Duguet:2025hwi} is included in the figure (dashed black line) for a qualitative comparison with our direct-only results, since its deformation parameter differs from $\beta_{20}$. Overall, the rigid-rotor and wave-function-rotor results nearly coincide. Both exhibit a nearly linear dependence on $\beta_{20}^2$, with close fit lines.

In extracting $\beta_{20}$ and $\langle\varepsilon_2^2\rangle$, we use the same values of $\kappa$ in the two constructions, yet both calculated quantities can differ. Their crossing locations need not coincide, because $\langle\varepsilon_2^2\rangle$ depends on laboratory one-body moments and two-body correlations as well as on $\kappa$.

First, for the independent-source baseline, a very weak linear drift behavior versus $\beta_{20}^2$ remains visible. The resulting slope stays far smaller than those of the full results dominated by $\langle \hat{\mathcal{E}}_{n} \rangle$. 

Second, retaining only the direct contribution of the two-body density (exchange contribution switched off),
both the rigid-rotor and wave-function-rotor results remain excellently linear: 
the fitted intercept stays close to the $1/A$ scale,
while the slope lies visibly above the Gaussian rigid-rotor benchmark 
$\langle\varepsilon_2^2\rangle=1/A+(3/4\pi)\beta_2^2$ 
(dashed line in Fig.~\ref{fig:eps2-vs-beta2surf}) \cite{Duguet:2025hwi}.
In that model the nuclear density is described by a Gaussian distribution with angle-dependent width, 
and $\beta_2$ is simply the parameter that controls this angular variation of the overall size. 
In contrast, $\beta_{20}$ in Fig.~\ref{fig:eps2-vs-beta2surf} is taken from an outer isodensity contour 
of the Nilsson density: it measures the shape of the nuclear surface, not the global width of a Gaussian distribution. 
The different deformation measures prevent a direct comparison of these fitted slopes as pair-correlation strengths.

Third, a complete calculation must include the exchange contribution. Restoring the Pauli term leaves the approximate linear trend intact, yet both the fitted intercept and slope become smaller than the direct-only results in the parameter space. The exchange term represents Pauli suppression for nucleons with the same spin-isospin labels. In this parameter space its contribution to $\langle\varepsilon_2^2\rangle$ grows in magnitude as $\kappa$ decreases. This trend follows from the weighted azimuthal pair moment in Eq.~\eqref{eq:eps-decomp}; local suppression alone does not determine it.

For outer contours with $f<0.2$ the quantitative conclusion is essentially unchanged; for $f\gtrsim0.3$, $\beta_{20}$ begins to mix in interior density differences, and the rigid-rotor and wave-function-rotor curves no longer coincide.
Since the result holds for these outer contours, the overlapping linear relation is stable within the selected model parameters. 
Within this controlled microscopic model, the $n=2$ pattern provides a useful benchmark
for relating geometric eccentricity to nuclear deformation in ultracentral heavy-ion collisions.

\begin{figure}[htbp]
\centering
\includegraphics[width=0.6\textwidth]{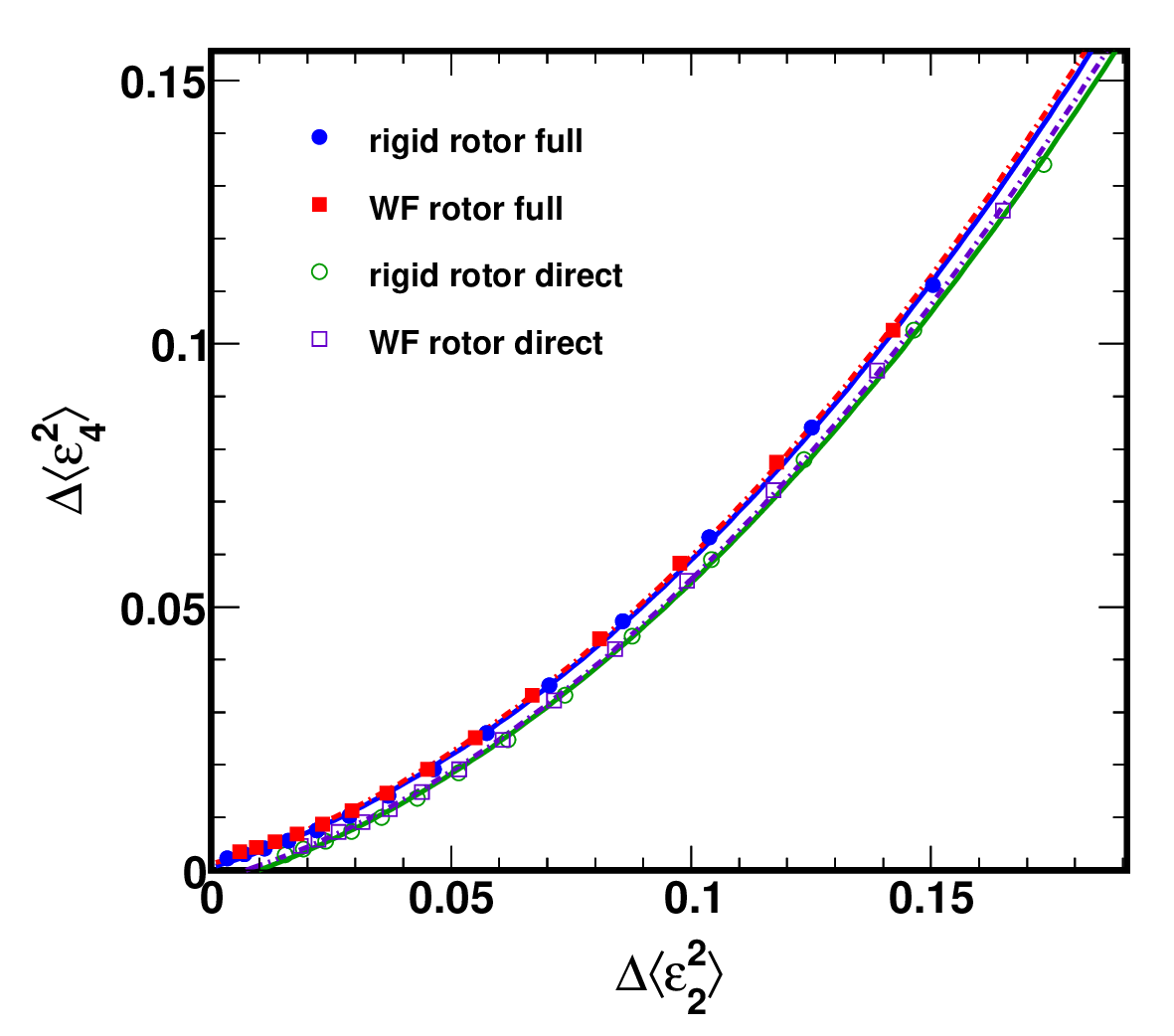}
\caption{Correlation-driven effective variance for Ne-20: 
$\Delta \langle\varepsilon_4^2\rangle = \langle\varepsilon_4^2\rangle-\langle\varepsilon_4^2\rangle_{\mathrm{indep}}$ versus
$\Delta \langle\varepsilon_2^2\rangle = \langle\varepsilon_2^2\rangle-\langle\varepsilon_2^2\rangle_{\mathrm{indep}}$.
Symbols for the direct-only and full results are same as in Fig.~\ref{fig:eps-vs-kappa}.
Solid and dash-dotted curves are quadratic fits of the rigid-rotor and wave-function-rotor results, respectively.}
\label{fig:eps-minus-indep-n2n4}
\end{figure}

We now turn to the results of hexadecapole variance with $n=4$ and discuss its relation to the quadrupole one. 
Because the independent-source baseline for $n=4$ component varies with deformation, the full variance  
$\langle\varepsilon_4^2\rangle$ still contains deformation dependence and is not a clean variable for testing the correlation response. 
Subtracting the independent-source term from the variance isolates $\langle\hat{\mathcal{E}}_{n}\rangle$ in the numerator, while the denominator still contains a one-body moment. Therefore Fig. \ref{fig:eps-minus-indep-n2n4} plots
the effective variance $\Delta \langle\varepsilon_4^2\rangle = \langle\varepsilon_4^2\rangle-\langle\varepsilon_4^2\rangle_{\mathrm{indep}}$ versus $\Delta \langle\varepsilon_2^2\rangle =  \langle\varepsilon_2^2\rangle-\langle\varepsilon_2^2\rangle_{\mathrm{indep}}$.

In the figure, the full and direct-only results form two roughly parallel but clearly separated approximate parabolas. The rigid-rotor and wave-function-rotor points lie close to one another on each trend. 
Pauli blocking effects reduce $\Delta \langle\varepsilon_2^2\rangle$ more than $\Delta \langle\varepsilon_4^2\rangle$, so the full result curve lies to the left side of the direct-only curve.

As shown in Fig.~\ref{fig:eps-minus-indep-n2n4}, both axes are calculated initial-geometry quantities. Their near overlap is an empirical feature of the Nilsson model in the current parameter space. The link between $\langle\hat{\mathcal{E}}_{4}\rangle$ and $\langle\hat{\mathcal{E}}_{2}\rangle$ deserves closer study to identify quantities more sensitive to the rotational construction; connecting them to measured flow requires a collision calculation.

%====================================================================
\section{Summary and Outlook}
\label{sec:summary}

We compare two orientational averages that generate the laboratory one- and two-body densities of Ne-20 nucleus in an analytically solvable model for deformed nuclei. In the rigid rotor, the orientational average is taken for the intrinsic probability density, while in the wave-function rotor, the average is taken for the many-body wave function. Both cases were constructed from a wave function in the form of Slater determinants in an axially deformed harmonic-oscillator model, with the oscillator ratio $\kappa$ as the only deformation parameter.

The central comparison is made at fixed $\kappa$ for both rotors. The wave-function-rotor outer contours generally contract along the symmetry axis, the equatorial ring is reduced, and the contour coefficients $\beta_{20}$ and $\beta_{40}$ are smaller than the rigid-rotor results in most parameter space. The calculated eccentricity variances $\langle\varepsilon_2^2\rangle$ and $\langle\varepsilon_4^2\rangle$ also differ between two rotors. They generally decrease as $\kappa$ approaches unity, and at given $\kappa$ the wave-function-rotor results often lie below the rigid-rotor ones. Their differences receive contributions from both laboratory one-body moments and two-body pair harmonics. Pauli blocking effects lower the eccentricity in both rotors without changing the ordering. 

We find simple empirical relations among the calculated contour and eccentricity quantities. 
The contour relation $\beta_{40}=p+q\beta_{20}^2$ has similar fitted slopes in both rotors. 
The quadrupole variance $\langle\varepsilon_2^2\rangle$ is approximately linear in $\beta_{20}^2$ in most parameter space. 
The relations between the hexadecapole baseline-subtracted (or effective) variance $\Delta\langle\varepsilon_4^2\rangle$ and the quadrupole one $\Delta\langle\varepsilon_2^2\rangle$ follow almost the same quadratic function for both rotors. These relations characterize the selected Nilsson model and contour prescription. 

Our calculation is based on the Nilsson model for deformed nuclei which can be solved analytically. Natural extensions include more realistic interactions and heavier nuclei, implementing laboratory densities into a full Glauber or hydrodynamic simulation, and comparison with measured flows in $^{16}$O+$^{16}$O and $^{20}$Ne+$^{20}$Ne collisions \cite{ALICE:2025luc}. 

%====================================================================

\textit{Acknowledgement}. The authors are supported by the National Natural Science Foundation of China under Grant Numbers 12575148, 12675172 and 12135011. 

\bibliographystyle{unsrt}
\bibliography{ref-nilsson}

@article{STAR:2024wgy,
    author = "Abdulhamid, M. I. and others",
    collaboration = "STAR",
    title = "{Imaging shapes of atomic nuclei in high-energy nuclear collisions}",
    eprint = "2401.06625",
    archivePrefix = "arXiv",
    primaryClass = "nucl-ex",
    doi = "10.1038/s41586-024-08097-2",
    journal = "Nature",
    volume = "635",
    number = "8037",
    pages = "67--72",
    year = "2024"
}

@article{Miller:2007ri,
    author = "Miller, Michael L. and Reygers, Klaus and Sanders, Stephen J. and Steinberg, Peter",
    title = "{Glauber modeling in high energy nuclear collisions}",
    eprint = "nucl-ex/0701025",
    archivePrefix = "arXiv",
    doi = "10.1146/annurev.nucl.57.090506.123020",
    journal = "Ann. Rev. Nucl. Part. Sci.",
    volume = "57",
    pages = "205--243",
    year = "2007"
}

@article{Xu:2026llz,
    author = "Xu, Hao-jie and Wang, Qun",
    title = "{Quantum Symmetry Restoration and Emergent Effective Deformation in Relativistic Heavy-Ion Collisions}",    	
    note = "arXiv:2606.02412 [nucl-th]",
    eprint = "2606.02412",
    archivePrefix = "arXiv",
    primaryClass = "nucl-th"
}

@article{Alvioli:2009ab,
    author = "Alvioli, M. and Drescher, H. -J. and Strikman, M.",
    title = "{A Monte Carlo generator of nucleon configurations in complex nuclei including Nucleon-Nucleon correlations}",
    eprint = "0905.2670",
    archivePrefix = "arXiv",
    primaryClass = "nucl-th",
    doi = "10.1016/j.physletb.2009.08.067",
    journal = "Phys. Lett. B",
    volume = "680",
    pages = "225--230",
    year = "2009"
}

@article{Broniowski:2010jd,
    author = "Broniowski, Wojciech and Rybczynski, Maciej",
    title = "{Two-body nucleon-nucleon correlations in Glauber models of relativistic heavy-ion collisions}",
    eprint = "1003.1088",
    archivePrefix = "arXiv",
    primaryClass = "nucl-th",
    doi = "10.1103/PhysRevC.81.064909",
    journal = "Phys. Rev. C",
    volume = "81",
    pages = "064909",
    year = "2010"
}

@article{Blaizot:2014wba,
    author = "Blaizot, Jean-Paul and Broniowski, Wojciech and Ollitrault, Jean-Yves",
    title = "{Correlations in the Monte Carlo Glauber model}",
    eprint = "1405.3274",
    archivePrefix = "arXiv",
    primaryClass = "nucl-th",
    reportNumber = "SACLAY-T14-060",
    doi = "10.1103/PhysRevC.90.034906",
    journal = "Phys. Rev. C",
    volume = "90",
    number = "3",
    pages = "034906",
    year = "2014"
}

@article{Duguet:2025hwi,
    author = "Duguet, Thomas and Giacalone, Giuliano and Jeon, Sangyong and Tichai, Alexander",
    title = "{Revealing the Harmonic Structure of Nuclear Two-Body Correlations in High-Energy Heavy-Ion Collisions}",
    eprint = "2504.02481",
    archivePrefix = "arXiv",
    primaryClass = "nucl-th",
    reportNumber = "CERN-TH-2025-056",
    doi = "10.1103/v2z7-wlnr",
    journal = "Phys. Rev. Lett.",
    volume = "135",
    number = "18",
    pages = "182301",
    year = "2025"
}

@article{Lowdin:1955zzb,
    author = "Lowdin, Per-Olov",
    title = "{Quantum Theory of Many-Particle Systems. 1. Physical Interpretations by Means of Density Matrices, Natural Spin-Orbitals, and Convergence Problems in the Method of Configurational Interaction}",
    doi = "10.1103/PhysRev.97.1474",
    journal = "Phys. Rev.",
    volume = "97",
    pages = "1474--1489",
    year = "1955"
}

@article{Lowdin:1955zza,
    author = "Lowdin, Per-Olov",
    title = "{Quantum Theory of Many-Particle Systems. 2. Study of the Ordinary Hartree-Fock Approximation}",
    doi = "10.1103/PhysRev.97.1490",
    journal = "Phys. Rev.",
    volume = "97",
    pages = "1490--1508",
    year = "1955"
}

@article{Giacalone:2023hwk,
    author = "Giacalone, Giuliano",
    title = "{Many-body correlations for nuclear physics across scales: from nuclei to quark-gluon plasmas to hadron distributions}",
    eprint = "2305.19843",
    archivePrefix = "arXiv",
    primaryClass = "nucl-th",
    doi = "10.1140/epja/s10050-023-01200-7",
    journal = "Eur. Phys. J. A",
    volume = "59",
    pages = "297",
    year = "2023"
}

@article{Blaizot:2025scr,
    author = "Blaizot, Jean-Paul and Giacalone, Giuliano",
    title = "{Angular structure of many-body correlations in atomic nuclei}",
    eprint = "2504.15421",
    archivePrefix = "arXiv",
    primaryClass = "nucl-th",
    doi = "10.1140/epja/s10050-025-01679-2",
    journal = "Eur. Phys. J. A",
    volume = "61",
    pages = "220",
    year = "2025"
}

@article{Ke:2025tyv,
    author = "Ke, Weiyao",
    title = "{From wave-function to fireball geometry: the role of a restored broken symmetry in ultra-relativistic collisions of deformed nuclei}",
    note = "arXiv:2509.09549 [nucl-th]",
    eprint = "2509.09549",
    archivePrefix = "arXiv",
    primaryClass = "nucl-th"
}

@article{Bofos:2026huw,
    author = "Bofos, Stavros and Bally, Benjamin and Duguet, Thomas and Frosini, Mikael",
    title = "{Imaging two-body correlations in atomic nuclei via low- and high-energy processes}",
    eprint = "2602.09890",
    archivePrefix = "arXiv",
    primaryClass = "nucl-th",
    doi = "10.1016/j.physletb.2026.140649",
    journal = "Phys. Lett. B",
    volume = "879",
    pages = "140649",
    year = "2026"
}

@article{Bofos:2026nmg,
    author = "Bofos, Stavros and Li, Yi and Ding, Chenrong and Bally, Benjamin and Duguet, Thomas and Frosini, Mikael and Yao, Jiangming",
    title = "{Quantum effects in the quadrupole rotor picture of ultra-relativistic ion-ion collisions}",
    note = "arXiv:2605.28813 [nucl-th]",
    eprint = "2605.28813",
    archivePrefix = "arXiv",
    primaryClass = "nucl-th"
}

@article{STAR:2021mii,
    author = "Abdallah, Mohamed and others",
    collaboration = "STAR",
    title = "{Search for the chiral magnetic effect with isobar collisions at $\sqrt{s_{NN}}=200$ GeV by the STAR Collaboration at RHIC}",
    eprint = "2109.00131",
    archivePrefix = "arXiv",
    primaryClass = "nucl-ex",
    doi = "10.1103/PhysRevC.105.014901",
    journal = "Phys. Rev. C",
    volume = "105",
    pages = "014901",
    year = "2022"
}

@article{STAR:2015mki,
    author = "Adamczyk, L. and others",
    collaboration = "STAR",
    title = "{Azimuthal anisotropy in U$+$U and Au$+$Au collisions at RHIC}",
    eprint = "1505.07812",
    archivePrefix = "arXiv",
    primaryClass = "nucl-ex",
    doi = "10.1103/PhysRevLett.115.222301",
    journal = "Phys. Rev. Lett.",
    volume = "115",
    pages = "222301",
    year = "2015"
}

@article{ALICE:2018lao,
    author = "Acharya, S. and others",
    collaboration = "ALICE",
    title = "{Anisotropic flow in Xe--Xe collisions at $\sqrt{s_{\rm NN}} = 5.44$ TeV}",
    eprint = "1805.01832",
    archivePrefix = "arXiv",
    primaryClass = "nucl-ex",
    doi = "10.1016/j.physletb.2018.06.059",
    journal = "Phys. Lett. B",
    volume = "784",
    pages = "82--95",
    year = "2018"
}

@article{ATLAS:2022dov,
    author = "Aad, G. and others",
    collaboration = "ATLAS",
    title = "{Correlations between flow and transverse momentum in Xe$+$Xe and Pb$+$Pb collisions at the LHC with the ATLAS detector}",
    eprint = "2205.00039",
    archivePrefix = "arXiv",
    primaryClass = "nucl-ex",
    doi = "10.1103/PhysRevC.107.054910",
    journal = "Phys. Rev. C",
    volume = "107",
    pages = "054910",
    year = "2023"
}

@article{CMS:2019cyz,
    author = "Sirunyan, A. M. and others",
    collaboration = "CMS",
    title = "{Charged-particle angular correlations in XeXe collisions at $\sqrt{s_{_{\mathrm{NN}}}} = 5.44$ TeV}",
    eprint = "1901.07997",
    archivePrefix = "arXiv",
    primaryClass = "hep-ex",
    doi = "10.1103/PhysRevC.100.044902",
    journal = "Phys. Rev. C",
    volume = "100",
    pages = "044902",
    year = "2019"
}

@article{ALICE:2025luc,
    author = "Acharya, Shreyasi and others",
    collaboration = "ALICE",
    title = "{Evidence of nuclear geometry-driven anisotropic flow in O$+$O and Ne$+$Ne collisions at $\sqrt{s_{NN}} = 5.36$ TeV}",
    eprint = "2509.06428",
    archivePrefix = "arXiv",
    primaryClass = "nucl-ex",
    doi = "10.1103/gymp-vp87",
    journal = "Phys. Rev. Lett.",
    volume = "137",
    pages = "082301",
    year = "2026"
}

@article{Giacalone:2019pca,
    author = "Giacalone, Giuliano",
    title = "{Observing the deformation of nuclei with relativistic nuclear collisions}",
    eprint = "1910.04673",
    archivePrefix = "arXiv",
    primaryClass = "nucl-th",
    doi = "10.1103/PhysRevLett.124.202301",
    journal = "Phys. Rev. Lett.",
    volume = "124",
    pages = "202301",
    year = "2020"
}

@article{Giacalone:2020awm,
    author = "Giacalone, Giuliano",
    title = "{Constraining the quadrupole deformation of atomic nuclei with relativistic nuclear collisions}",
    eprint = "2004.14463",
    archivePrefix = "arXiv",
    primaryClass = "nucl-th",
    doi = "10.1103/PhysRevC.102.024901",
    journal = "Phys. Rev. C",
    volume = "102",
    pages = "024901",
    year = "2020"
}

@article{Giacalone:2021udy,
    author = "Giacalone, Giuliano and Jia, Jiangyong and Zhang, Chunjian",
    title = "{Impact of nuclear deformation on relativistic heavy-ion collisions: Assessing consistency in nuclear physics across energy scales}",
    eprint = "2105.01638",
    archivePrefix = "arXiv",
    primaryClass = "nucl-th",
    doi = "10.1103/PhysRevLett.127.242301",
    journal = "Phys. Rev. Lett.",
    volume = "127",
    pages = "242301",
    year = "2021"
}

@article{Zhang:2021kxj,
    author = "Zhang, Chunjian and Jia, Jiangyong",
    title = "{Evidence of quadrupole and octupole deformations in $^{96}$Zr+$^{96}$Zr and $^{96}$Ru+$^{96}$Ru collisions at ultrarelativistic energies}",
    eprint = "2109.01631",
    archivePrefix = "arXiv",
    primaryClass = "nucl-th",
    doi = "10.1103/PhysRevLett.128.022301",
    journal = "Phys. Rev. Lett.",
    volume = "128",
    pages = "022301",
    year = "2022"
}

@article{Bally:2021qys,
    author = "Bally, Benjamin and Bender, Michael and Giacalone, Giuliano and Som\`a, Vittorio",
    title = "{Evidence of the triaxial structure of $^{129}$Xe at the Large Hadron Collider}",
    eprint = "2108.09578",
    archivePrefix = "arXiv",
    primaryClass = "nucl-th",
    doi = "10.1103/PhysRevLett.128.082301",
    journal = "Phys. Rev. Lett.",
    volume = "128",
    pages = "082301",
    year = "2022"
}

@article{Ryssens:2023fkv,
    author = "Ryssens, Wouter and Giacalone, Giuliano and Schenke, Bj{\"o}rn and Shen, Chun",
    title = "{Evidence of hexadecapole deformation in uranium-238 at the Relativistic Heavy Ion Collider}",
    eprint = "2302.13617",
    archivePrefix = "arXiv",
    primaryClass = "nucl-th",
    doi = "10.1103/PhysRevLett.130.212302",
    journal = "Phys. Rev. Lett.",
    volume = "130",
    pages = "212302",
    year = "2023"
}

@article{Xu:2024bdh,
    author = "Xu, Hao-jie and Zhao, Jie and Wang, Fuqiang",
    title = "{Hexadecapole deformation of $^{238}$U from relativistic heavy-ion collisions using a nonlinear response coefficient}",
    eprint = "2402.16550",
    archivePrefix = "arXiv",
    primaryClass = "nucl-th",
    doi = "10.1103/PhysRevLett.132.262301",
    journal = "Phys. Rev. Lett.",
    volume = "132",
    pages = "262301",
    year = "2024"
}

@article{Heinz:2013th,
    author = "Heinz, Ulrich and Snellings, Raimond",
    title = "{Collective flow and viscosity in relativistic heavy-ion collisions}",
    eprint = "1301.2826",
    archivePrefix = "arXiv",
    primaryClass = "nucl-th",
    doi = "10.1146/annurev-nucl-102212-170540",
    journal = "Ann. Rev. Nucl. Part. Sci.",
    volume = "63",
    pages = "123--151",
    year = "2013"
}

@book{Bohr:1975zz,
    author = "Bohr, Aage and Mottelson, Ben R.",
    title = "{Nuclear Structure, Vol.~II: Nuclear Deformations}",
    publisher = "W.~A.~Benjamin",
    address = "Reading, Massachusetts",
    year = "1975"
}

@article{Shou:2014eya,
    author = "Shou, Q. Y. and Ma, Y. G. and Sorensen, P. and Tang, A. H. and Videb{\ae}k, F. and Wang, H.",
    title = "{Parameterization of deformed nuclei for Glauber modeling in relativistic heavy ion collisions}",
    eprint = "1409.8375",
    archivePrefix = "arXiv",
    primaryClass = "nucl-th",
    doi = "10.1016/j.physletb.2015.07.078",
    journal = "Phys. Lett. B",
    volume = "749",
    pages = "215--220",
    year = "2015"
}

@article{Jia:2022ozr,
    author = "Jia, Jiangyong and others",
    title = "{Imaging the initial condition of heavy-ion collisions and nuclear structure across the nuclide chart}",
    eprint = "2209.11042",
    archivePrefix = "arXiv",
    primaryClass = "nucl-ex",
    doi = "10.1007/s41365-024-01589-w",
    journal = "Nucl. Sci. Tech.",
    volume = "35",
    pages = "220",
    year = "2024"
}

@book{Ring:1980,
    author = "Ring, Peter and Schuck, Peter",
    title = "{The Nuclear Many-Body Problem}",
    publisher = "Springer-Verlag",
    address = "New York",
    year = "1980"
}

@article{Lee:2025req,
    author = "Lee, Dean",
    title = "{Lattice effective field theory simulations of nuclei}",
    eprint = "2501.03303",
    archivePrefix = "arXiv",
    primaryClass = "nucl-th",
    doi = "10.1146/annurev-nucl-101918-023343",
    journal = "Ann. Rev. Nucl. Part. Sci.",
    volume = "75",
    pages = "109--128",
    year = "2025"
}

@article{Frosini:2021sxj,
    author = "Frosini, Mikael and Duguet, Thomas and Ebran, Jean-Paul and Bally, Benjamin and Mongelli, Tobias and Rodr{\'i}guez, Tom{\'a}s R. and Roth, Robert and Som{\`a}, Vittorio",
    title = "{Multi-reference many-body perturbation theory for nuclei II. Ab initio study of neon isotopes via PGCM and IM-NCSM calculations}",
    eprint = "2111.00797",
    archivePrefix = "arXiv",
    primaryClass = "nucl-th",
    doi = "10.1140/epja/s10050-022-00693-y",
    journal = "Eur. Phys. J. A",
    volume = "58",
    pages = "63",
    year = "2022"
}

@article{ATLAS:2025nnt,
    author = "Aad, G. and others",
    collaboration = "ATLAS",
    title = "{Measurement of the azimuthal anisotropy of charged particles in $\sqrt{s_{\rm NN}} = 5.36$ TeV $^{16}$O$+^{16}$O and $^{20}$Ne$+^{20}$Ne collisions with the ATLAS detector}",
    eprint = "2509.05171",
    archivePrefix = "arXiv",
    primaryClass = "nucl-ex",
    doi = "10.1103/xqxz-8bhf",
    journal = "Phys. Rev. C",
    volume = "113",
    pages = "045205",
    year = "2026"
}

@article{CMS:2025tga,
    author = "Hayrapetyan, A. and others",
    collaboration = "CMS",
    title = "{Observation of long-range collective flow in OO and NeNe collisions and implications for nuclear structure studies}",
    eprint = "2510.02580",
    archivePrefix = "arXiv",
    primaryClass = "nucl-ex",
    doi = "10.1103/26wx-tg6f",
    journal = "Phys. Rev. Lett.",
    volume = "137",
    pages = "082302",
    year = "2026"
}

@article{Blaizot:2025bfu,
    author = "Blaizot, Jean-Paul and Giacalone, Giuliano and Lovato, Alessandro",
    title = "{Nuclear collectivity and the harmonic spectrum of two-body correlations}",
    note = "arXiv:2512.18926 [nucl-th]",
    eprint = "2512.18926",
    archivePrefix = "arXiv",
    primaryClass = "nucl-th"
}

@article{PHENIX:2018lia,
    author = "Aidala, C. and others",
    collaboration = "PHENIX",
    title = "{Creation of quark--gluon plasma droplets with three distinct geometries}",
    eprint = "1805.02973",
    archivePrefix = "arXiv",
    primaryClass = "nucl-ex",
    doi = "10.1038/s41567-018-0360-0",
    journal = "Nature Phys.",
    volume = "15",
    pages = "214--220",
    year = "2019"
}

@article{Nagle:2018nvi,
    author = "Nagle, J. L. and Zajc, W. A.",
    title = "{Small system collectivity in relativistic hadronic and nuclear collisions}",
    eprint = "1801.03477",
    archivePrefix = "arXiv",
    primaryClass = "nucl-ex",
    doi = "10.1146/annurev-nucl-101916-123209",
    journal = "Ann. Rev. Nucl. Part. Sci.",
    volume = "68",
    pages = "211--235",
    year = "2018"
}

@article{Giacalone:2024luz,
    author = "Giacalone, Giuliano and others",
    title = "{Exploiting $^{20}$Ne isotopes for precision characterizations of collectivity in small systems}",
    eprint = "2402.05995",
    archivePrefix = "arXiv",
    primaryClass = "nucl-th",
    doi = "10.1103/k8rb-jgvq",
    journal = "Phys. Rev. Lett.",
    volume = "135",
    pages = "012302",
    year = "2025"
}

@article{Li:2019kkh,
    author = "Li, Hanlin and Xu, Hao-jie and Zhou, Ying and Wang, Xiaobao and Zhao, Jie and Chen, Lie-Wen and Wang, Fuqiang",
    title = "{Probing the neutron skin with ultrarelativistic isobaric collisions}",
    eprint = "1910.06170",
    archivePrefix = "arXiv",
    primaryClass = "nucl-th",
    doi = "10.1103/PhysRevLett.125.222301",
    journal = "Phys. Rev. Lett.",
    volume = "125",
    pages = "222301",
    year = "2020"
}

@article{Jia:2022qgl,
    author = "Jia, Jiangyong and Giacalone, Giuliano and Zhang, Chunjian",
    title = "{Separating the impact of nuclear skin and nuclear deformation in high-energy isobar collisions}",
    eprint = "2206.10449",
    archivePrefix = "arXiv",
    primaryClass = "nucl-th",
    doi = "10.1103/PhysRevLett.131.022301",
    journal = "Phys. Rev. Lett.",
    volume = "131",
    pages = "022301",
    year = "2023"
}

@article{Jia:2021wbq,
    author = "Jia, Jiangyong and Huang, Shengli and Zhang, Chunjian",
    title = "{Probing nuclear quadrupole deformation from correlation of elliptic flow and transverse momentum in heavy ion collisions}",
    eprint = "2105.05713",
    archivePrefix = "arXiv",
    primaryClass = "nucl-th",
    doi = "10.1103/PhysRevC.105.014906",
    journal = "Phys. Rev. C",
    volume = "105",
    pages = "014906",
    year = "2022"
}

@article{Song:2017wtw,
    author = "Song, Huichao and Zhou, You and Gajdosova, Katarina",
    title = "{Collective flow and hydrodynamics in large and small systems at the LHC}",
    eprint = "1703.00670",
    archivePrefix = "arXiv",
    primaryClass = "nucl-th",
    doi = "10.1007/s41365-017-0245-4",
    journal = "Nucl. Sci. Tech.",
    volume = "28",
    pages = "99",
    year = "2017"
}

@article{Zhao:2017rgg,
    author = "Zhao, Wenbin and Zhou, You and Xu, Hao-jie and Deng, Weitian and Song, Huichao",
    title = "{Hydrodynamic collectivity in proton--proton collisions at 13 TeV}",
    eprint = "1801.00271",
    archivePrefix = "arXiv",
    primaryClass = "nucl-th",
    doi = "10.1016/j.physletb.2018.03.022",
    journal = "Phys. Lett. B",
    volume = "780",
    pages = "495--500",
    year = "2018"
}

@article{Zhao:2020wcd,
    author = "Zhao, Wenbin and Ko, Che Ming and Liu, Yu-Xin and Qin, Guang-You and Song, Huichao",
    title = "{Probing the partonic degrees of freedom in high multiplicity p--Pb collisions at $\sqrt{s_{NN}}=5.02$ TeV}",
    eprint = "1911.00826",
    archivePrefix = "arXiv",
    primaryClass = "nucl-th",
    doi = "10.1103/PhysRevLett.125.072301",
    journal = "Phys. Rev. Lett.",
    volume = "125",
    pages = "072301",
    year = "2020"
}

@article{YuanyuanWang:2024sgp,
    author = "Wang, Yuanyuan and Zhao, Shujun and Cao, Boxing and Xu, Hao-jie and Song, Huichao",
    title = "{Exploring the compactness of $\alpha$ cluster in $^{16}$O nuclei with relativistic $^{16}$O+$^{16}$O collisions}",
    eprint = "2401.15723",
    archivePrefix = "arXiv",
    primaryClass = "nucl-th",
    doi = "10.1103/PhysRevC.109.L051904",
    journal = "Phys. Rev. C",
    volume = "109",
    pages = "L051904",
    year = "2024"
}

@article{Hu:2025eid,
    author = "Hu, Jin-Yu and Xu, Hao-jie and Wang, Xiaobao and Pu, Shi",
    title = "{Probing the tetrahedral $\alpha$ clusters in relativistic $^{16}$O+$^{16}$O collisions}",
    eprint = "2507.01493",
    archivePrefix = "arXiv",
    primaryClass = "nucl-th",
    doi = "10.1103/z96t-6rxd",
    journal = "Phys. Rev. C",
    volume = "114",
    pages = "034905",
    year = "2026"
}

@article{Zhao:2024feh,
    author = "Zhao, Xin-Li and Zhou, You and Lin, Zi-Wei and Zhang, Chao and Ma, Guo-Liang",
    title = "{Nuclear cluster structure effect in $^{16}$O+$^{16}$O collisions at the top RHIC energy}",
    eprint = "2404.09780",
    archivePrefix = "arXiv",
    primaryClass = "nucl-th",
    doi = "10.1016/j.physletb.2026.140254",
    journal = "Phys. Lett. B",
    volume = "874",
    pages = "140254",
    year = "2026"
}

@article{Liu:2025zsi,
    author = "Liu, Lu-Meng and Wang, Hai-Cheng and Li, Song-Jie and Zhang, Chunjian and Xu, Jun and Ren, Zhong-Zhou and Jia, Jiangyong and Huang, Xu-Guang",
    title = "{Directly probing existence of $\alpha$-cluster structure in $^{20}$Ne by relativistic heavy-ion collisions}",
    eprint = "2502.08057",
    archivePrefix = "arXiv",
    primaryClass = "nucl-th",
    doi = "10.1103/PhysRevC.111.L021901",
    journal = "Phys. Rev. C",
    volume = "111",
    pages = "L021901",
    year = "2025"
}

@article{Li:2025hae,
    author = "Li, Pei and Zhou, Bo and Ma, Guo-Liang",
    title = "{Identifying $\alpha$-cluster configurations in $^{20}$Ne via ultracentral Ne+Ne collisions}",
    eprint = "2504.04688",
    archivePrefix = "arXiv",
    primaryClass = "nucl-th",
    doi = "10.1103/tffz-8q1m",
    journal = "Phys. Rev. Lett.",
    volume = "136",
    pages = "082302",
    year = "2026"
}

\end{document}